\documentclass[prd,aps,twocolumn,nofootinbib,superscriptaddress,eqsecnum,floatfix,preprintnumbers,amsmath,amssymb,
nofootinbib,longbibliography]{revtex4-1}

\usepackage{amssymb,amsmath}
\usepackage{epsfig}
\usepackage[dvipsnames]{xcolor}
\usepackage[utf8]{inputenc}
\usepackage{stmaryrd}
\usepackage{mathrsfs}
\usepackage{mathalfa}
\usepackage{accents}
\usepackage[normalem]{ulem}
\usepackage{enumitem}

\usepackage{graphicx}
\graphicspath{ {c:/Documents/PhysicsNotes/GradResearch/Images/ChiralAnomaly/} }

\newcommand{\pa}{\partial}

\newcommand{\be}{\begin{equation}}
\newcommand{\ee}{\end{equation}}
\newcommand{\bea}{\begin{eqnarray}}
\newcommand{\eea}{\end{eqnarray}}
\newcommand{\ba}{\begin{equation}\begin{aligned}}
\newcommand{\ea}{\end{aligned}\end{equation}}

\newcommand{\beg}{\begin{gather*}}
\newcommand{\eng}{\end{gather*}}
\newcommand{\hh}{,\hspace{0.5cm}}
\newcommand{\hhh}{,\hspace{0.2cm}}

\newcommand{\n}[1]{\label{#1}}

\newcommand{\CAL}{\mathcal}

\newcommand{\ins}[1]{{\mbox{\tiny #1}}}

\newcommand{\inds}[1]{{\scriptscriptstyle #1}}

\newcommand{\ts}[1]{{\boldsymbol{#1}}}

\def\XXint#1#2#3{{\setbox0=\hbox{$#1{#2#3}{\int}$ }
\vcenter{\hbox{$#2#3$ }}\kern-.6\wd0}}

\usepackage[makeroom]{cancel}
\usepackage[caption=false]{subfig}
\usepackage[colorlinks=true,
            citecolor=green,
            linkcolor=red,
            filecolor=cyan,
            urlcolor=magenta,
            backref=false]{hyperref}

\newcommand{\val}[1]{{\color{red}{#1}}}

\newcommand{\bbi}[1]{\boldsymbol{{\bar{#1}}}}
\newcommand{\bi}[1]{{ \boldsymbol{{#1}}}}

\newcommand{\xx}[1]{\overset{\inds{#1}}{\xi}}

\newcommand{\sqg}{\sqrt{g}}

\newcommand{\sqq}{\sqrt{q}}
\newcommand{\sqf}{\sqrt{\gamma}}

\begin{document}

\title{Chiral anomalies in black hole spacetimes}
\author{Valeri P. Frolov}%
\email[]{vfrolov@ualberta.ca}
\affiliation{Theoretical Physics Institute, Department of Physics,
University of Alberta,\\
Edmonton, Alberta, T6G 2E1, Canada
}
\author{Alex Koek}
\email[]{koek@ualberta.ca}
\affiliation{Theoretical Physics Institute, Department of Physics,
University of Alberta,\\
Edmonton, Alberta, T6G 2E1, Canada
}
\author{Jose Pinedo Soto}
\email[]{pinedoso@ualberta.ca}
\affiliation{Theoretical Physics Institute, Department of Physics,
University of Alberta,\\
Edmonton, Alberta, T6G 2E1, Canada
}
\author{Andrei Zelnikov}%
\email[]{zelnikov@ualberta.ca}
\affiliation{Theoretical Physics Institute, Department of Physics,
University of Alberta,\\
Edmonton, Alberta, T6G 2E1, Canada
}


\begin{abstract}
We study the properties of chiral anomalies in a wide class of spacetimes which possess a principal Killing--Yano tensor. This class includes metrics of charged rotating black holes as a special physically important case. The spacetimes which admit a principal Killing-Yano tensor possess a number of remarkable properties. In particular, such spacetimes have two commuting Killing vectors and a Killing tensor responsible for their hidden symmetries. We calculate the gravitational and electromagnetic contributions to the axial anomaly currents in the spacetime of a charged rotating black hole, and demonstrate that the equation for the chiral anomaly current has special solutions which respect both explicit and hidden symmetries. Two of these solutions have the form of currents propagating along two principal null directions, which are null eigenvectors of the Riemann tensor. These solutions describe chiral currents for the incoming and outgoing polarization fluxes. It is demonstrated that these principal chiral currents can be written explicitly in the form which contains the off-shell metric coefficients and their derivatives. We discuss conditions where the principle chiral anomaly current is regular at the horizon and the axes of symmetry. We demonstrate that for states where the current vanishes at the past horizon and at the past null infinity, there exist chirality fluxes at both the future horizon and future infinity. The latter is directly related to the polarization asymmetry of Hawking radiation for massless spinning particles. We also calculate the Chern--Simons currents for both gravitational and electromagnetic chiral anomalies in the black hole spacetime, and discuss the properties of the chirality fluxes associated with these currents.
\end{abstract}

{ \hfill    Alberta Thy 11-22}

\maketitle

\section{Introduction}

There exists a well known correspondence between gravitation and electromagnetism. This correspondence becomes quite transparent when a linearized version of Einstein gravity is considered (a comprehensive review of this subject can be found e.g. in \cite{Mashhoon:2003}). For example, the gravitational interaction of a spin with a gravitational field is similar to the interaction of a magnetic dipole with an electromagnetic field. Both the spinning particle in a gravitational field of a massive rotating object, and the magnetic dipole in a static magnetic field undergo precession.
The equations describing the motion of massive spinning objects in a gravitational field are known as the Mathisson–Papapetrou–Dixon equations \cite{Mathisson:1937,Papapetrou:1951pa,Dixon}. Using this equation, Wald \cite{WALD:1972} evaluated the force on a spinning test body at rest in the exterior field of an arbitrary stationary, rotating source. In particular, he demonstrated that this force acting on a particle with spin $\vec{s}$ in the gravitational field of a massive object with angular momentum $\vec{L}$ depends on the relative orientation of vectors  $\vec{s}$ and $\vec{L}$.  For example, for a spinning particle on the symmetry axis of a rotating black hole, this force is repulsive if vectors $\vec{s}$ and $\vec{L}$ are parallel, and attractive if the vectors are anti-parallel.

When a spinning particle is brought into the vicinity of a spinning black hole along the black hole's symmetry axis, the energy of a particle with an anti-parallel spin orientation will be larger than that of a particle with a parallel spin orientation. The additional dependence of this energy on spin makes the quantum creation of particles with spin directed along the black hole's angular momentum more favorable. For the Hawking quantum flux of neutrinos, anti-neutrinos are predominantly are emitted in the direction of $\vec{L}$, while neutrinos are mainly emitted in the opposite direction \cite{Leahy:1979,Vilenkin:1979,Bolashenko:1989,Casals:2012es}. This effect has an analogue in the quantum radiation of photons and gravitons by a rotating black hole. Namely, there exists an asymmetry in the emission of left- and right-hand-polarized quanta of these fields in a given direction. As a result, the electromagnetic and gravitational quantum radiation of rotating black holes are polarized \cite{Bolashenko:1989trudy,Bolashenko:1989}. A similar effect is expected for higher-dimensional rotating black holes \cite{Casals_2009}.
This effect, as well as the connected property of asymmetry in Hawking radiation, might be important for the search of a decay of mini black holes if such black holes could be produced in colliders.

Dolgov and collaborators made an interesting observation. They pointed out that the effect of asymmetrical radiation of massless spinning particles by a rotating black hole is related to a so-called ``gravitational chiral anomaly" \cite{DOLGOV:1988,Dolgov:1989a,DOLGOV:1989b}.

An anomaly in quantum field theory is a well-known phenomenon that arises from a conflict between symmetries of the classical theory and its quantization (for a general discussion of this subject see e.g. \cite{Bertlmann:2000,Fujikawa2004path} and references therein).
A chiral anomaly occurs when a chiral current is not conserved in quantum theory, in spite of its classical conservation\footnote{
Similar effects of anomalous transport phenomena in chiral liquids \cite{Zakharov:2012vv,Chernodub:2021nff} are of great interest in condensed matter physics.}.
One of the most ``dramatic" consequences of the axial anomaly is the possible production of fermions whose quantum numbers violate classical laws.
An axial-current anomaly for the massless Dirac field $\psi$ with electric charge $e$ in external electromagnetic and gravitational fields was calculated in papers \cite{Kimura:1969,SALAM,EGUCHI,Prokhorov:2022rna}. Namely, if $a^{\mu}=\bar{\psi}\gamma^{\mu}\gamma^5 \psi$ is the axial current, then the quantum average of its divergence does not vanish. It has the form\footnote{\n{FOOT}
In this relation we use the sign convention of \cite{Misner:1973prb}.
The coefficient of the electromagnetic field contribution depends on the choice of units. In Heaviside units, this coefficient is $e^2/(2\pi)$ (see e.g. \cite{CAMBRIDGE}).
Let us also note that for a Weyl neutrino, the coefficient of the gravitational anomaly is twice as small, and is $1/(384\pi^2)$.
}

\be \n{FERM}
\langle \nabla_{\mu} a^{\mu}\rangle =\frac{e^2}{ 8\pi^2} F_{\mu\nu}{}^*\!F^{\mu\nu}-
\frac{1}{192\pi^2}R_{\mu\nu\alpha\beta}{}^*\!R^{\mu\nu\alpha\beta}\, ,
\ee
where
\ba
&{}^*\!F^{\mu\nu}=\frac{1}{ 2}e^{\mu\nu\alpha\beta}F_{\alpha\beta}\, ,\\
&{}^*\!R^{\mu\nu\alpha\beta}=\frac{1}{ 2}e^{\mu\nu\kappa\lambda}R_{\kappa\lambda}^{\ \ \alpha\beta}\, .
\ea

Dolgov and collaborators demonstrated that a similar chiral anomaly exists for an Abelian vector field $A_{\mu}$ \cite{Dolgov:1987yp,DOLGOV:1988,Dolgov:1989a,DOLGOV:1989b,Agullo:2018nfv}. Namely, they considered the current
\be
K^{\mu}=e^{\mu\nu\alpha\beta}A_{\nu}\pa_{\alpha}A_{\beta}\, ,
\ee
and showed that
\be
\langle \nabla_{\mu} K^{\mu}\rangle=-\frac{1}{ 96\pi^2}R_{\mu\nu\alpha\beta}{}^*\!R^{\mu\nu\alpha\beta}\, .
\ee
An approximate evaluation of the chiral current for the Riemann curvature tensor in the Kerr metric was performed in
\cite{DOLGOV:1988,Galaverni:2021}.
The physical interpretation of the currents $a^{\mu}$ and $K^{\mu}$ comes from the observation that integral over all space for components $a^0$ and $K^0$  has the dimensionality of an angular momentum, and is proportional to the difference between the right and left circularly-polarized components,  i.e. the net helicity \cite{Agullo:2018nfv}. Chiral anomalies for other fields and further references can be found in \cite{Prokhorov:2022rna}.
Calculations of chiral anomalies in four dimensions and higher can be found in the book \cite{nieu:2006}.

There exist a number of publications devoted to the study of chiral and other anomalies in black hole spacetimes.
A long time ago, Christensen and Fulling \cite{Christensen:1977jc} discussed conformal anomalies for a conformal massless field in the Schwarzschild geometry. For such a theory, the classical trace of the stress-energy tensor vanishes, while its quantum average does not. The authors showed that in two dimensions, the conformal anomaly uniquely determines the stress-energy tensor itself, up to two arbitrary functions of one variables that depend on the choice of state. In particular, for the Unruh vacuum, the flux of Hawking radiation at infinity is uniquely determined by the conformal anomaly. In four dimensions for a similar state, the expression for the quantum average of the stress-energy tensor, in addition to the known conformal anomaly, contains an unknown arbitrary function of one variable.

An interesting approach in connecting Hawking radiation from black holes to quantum anomalies was proposed and developed in the papers \cite{Robinson:2005pd,Iso:2006wa,Iso:2006ut,Banerjee:2007qs}. In this approach, one considers a free quantum field in the black hole background and focuses on its properties in a narrow strip near the horizon. After decomposition of the four-dimensional field into modes, one reduces the problem to an infinite set of two-dimensional fields with an effective potential that vanishes in the near-horizon domain. In this regime, one effectively has a free two-dimensional massless field with two kinds of modes: ``left movers" describing particles moving towards the horizon, and ``right movers" describing particles propagating towards the black hole exterior. By studying the gravitational anomalies in such a two-dimensional chiral model, Robinson and Wilczek \cite{Robinson:2005pd} demonstrated that the existence of the Hawking radiation is necessary for cancellation of this anomaly. The papers \cite{Iso:2006wa,Iso:2006ut,Banerjee:2007qs} contain a generalization of these results to charged and rotating black holes\footnote{
Let us note that this approach is similar to the case where one interprets the production of charged massless fermions by a one-dimensional electric potential step-function as an anomalous fermion production}\cite{Blaer:1981ps}.

Let us emphasize  that the above approach only allows one to obtain information about the quantum average of the stress-energy tensor of the field in the region very close to the horizon. In order to find the Hawking radiation at infinity, one needs to first figure out how each mode's contribution to the stress-energy tensor is affected by the effective potential in the domain where the adopted approximation is not valid, and then one needs to sum over all modes.

In the present paper, we follow the basic ideas of Dolgov and collaborators \cite{DOLGOV:1988,Dolgov:1989a,DOLGOV:1989b}, and we study the contribution of chiral anomalies to the asymmetry in net helicity of Hawking radiation from massless spinning particles.
Namely, we consider the equation
\be \n{MAIN}
\nabla_{\mu}J^{\mu}=\mathcal{P}\hh\,
\ee
for two cases: $ \mathcal{P}=-\frac{1}{2}R_{\mu\nu\alpha\beta}{}^*\!R^{\mu\nu\alpha\beta}$ and  $\mathcal{P}=\mathcal{P}_e=F_{\mu\nu}{}^*\!F^{\mu\nu}$.
Currents $\langle  a^{\mu}\rangle$ and $\langle K^{\mu}\rangle$ can be obtained as a linear combinations of the currents $J^{\mu}$ with an appropriate choice of the constant coefficients. We study solutions of these equations in a wide class of spacetimes, including the case of charged rotating black holes. This class is singled out by the property that they admit a special object called principal Killing-Yano tensor, which is a generator of hidden symmetries \cite{Frolov:2017kze,FrolovZelnikov:2011}. The principal Killing-Yano tensor determines a preferable (Darboux) frame and a system of canonical coordinates. Such a metric admits two commuting Killing vectors.
When written in canonical coordinates, this spacetime metric contains two arbitrary single-variable functions. If the Ricci scalar vanishes, these functions are second order polynomials of their arguments. The Kerr and Kerr-Newman metrics, describing a rotating black hole in an asymptotically flat spacetime, belong to this subclass of metrics.

We impose natural condition on the current $J^{\mu}$. Namely, we assume that it respects the spacetime symmetries and is regular at the horizon and symmetry axes. In order to incorporate these properties, we use a formalism developed by Geroch \cite{Geroch_1,Geroch_2}. He demonstrated that in the presence of two commuting Killing vectors, one can introduce a two-dimensional space $S$ the points of which are orbits of the Killing vectors. In other words, our spacetime can be decomposed into a bundle of such 2D Killing vector surfaces over $S$.
In such a description, vectors and tensors respecting the spacetime symmetries can be identified with corresponding 2D fields on $S$. One can use such a representation to reduce 4D equations to 2D equations on $S$. We apply this method to the spacetime of an eternal black hole. In this case, we impose an additional condition that there are no initial fluxes of the polarization, and we require that the current $J^{\mu}$ vanishes both at the past null infinity ${\cal J}^-$, and at the past horizon.

We describe a method of obtaining exact solutions to (\ref{MAIN}) that satisfy the imposed symmetry and regularity conditions. Such a solution is found in explicit analytical form. We use these results to obtain an analytic expression for the flux of the current at the infinity. In the second part of the paper, we repeat this analysis for the chiral anomaly generated by the $\ts{F}{}^*\!\ts{F}$ term in (\ref{FERM}) in the background of a charged rotating black hole.

It should be emphasized that there exists a freedom in the choice of a solution for (\ref{MAIN}). Namely, one can add a solution of the homogeneous equation, i.e. a solution of (\ref{MAIN}), with $\mathcal{P}=0$. We analyse this freedom, and discuss constraints imposed on it by the regularity conditions on the horizon and symmetry axes.

The paper is organized as follows. In Section \ref{s2} we remind the reader of the main points from Geroch's approach, and collect some useful formulas related with this formalism. In Section \ref{s3} we describe the main properties of spacetimes which admit a principal Killing-Yano tensor, and specify regularity conditions at both the horizon and the symmetry axes in such spacetimes. In Section \ref{s4} we study solutions of the equation for the chiral anomaly in the off-shell metric. We demonstrate that there exist solutions for the current which respect both explicit and hidden symmetries of the spacetime, and we find their explicit form. In this section we also describe principal chiral currents which are collinear with the principal null directions of the spacetime. In Section \ref{s5} we study the Chern-Simons form of the chiral current, and discuss its relation to the principal chiral currents. Section \ref{s6} is devoted to studying the contribution of the electromagnetic field to the chiral current anomaly in a spacetime of charged rotating black hole. Section \ref{s7} contains discussion of the obtained results.

In this paper we use natural units in which $G=c=\hbar=1$, and the signature conventions of the book \cite{Misner:1973prb}.

\section{Basic equations}
\n{s2}
\subsection{Geroch's formalism}

Let $M$ be a four-dimensional spacetime with a metric $g_{\mu\nu}$, which admits two commuting linearly-independent Killing vectors
$\xx{0}$ and $\xx{1}$.  Geroch \cite{Geroch_2} demonstrated that if additional natural conditions are imposed on these Killing vector fields (which we also assume to be valid), there exists a projection $\psi$ of the spacetime $M$ onto a smooth two-dimensional space $S$.

We denote
\ba
&\lambda_{00}={\xx{0}}{}^{\mu}\xx{0}_{\mu}\hh \lambda_{11}={\xx{1}}{}^{\mu}\xx{1}_{\mu}\hh
\lambda_{01}={\xx{0}}{}^{\mu}\xx{1}_{\mu}\, ,\\
&\gamma_{\mu\nu}=\val-\frac{1}{ \gamma}\Big[
\lambda_{11}\xx{0}_{\mu}\xx{0}_{\nu}+\lambda_{00}\xx{1}_{\mu}\xx{1}_{\nu}-2\lambda_{01}\xx{0}_{(\mu}\xx{1}_{\nu)}
\Big]\, ,\\
&\rho_{\mu\nu}=2\xx{0}_{[\mu}\xx{1}_{\nu]}\, .
\ea

In what follows we also assume that the following so-called ``circularity conditions" are satisfied (see e.g. \cite{kramer:2003})
\be \n{CIRC}
e^{\alpha\beta\gamma\delta} {\xx{0}}{}_{\alpha}{\xx{1}}{}_{\beta}{\xx{0}}{}_{\gamma;\delta}=0\hh e^{\alpha\beta\gamma\delta} {\xx{0}}{}_{\alpha}{\xx{1}}{}_{\beta}{\xx{1}}{}_{\gamma;\delta}=0\, .
\ee
These relations are necessary and sufficient conditions for the 2-flats orthogonal to $\xx{0}$ and $\xx{1}$ to be integrable. Let us denote by $\Gamma$ the two-dimensional span of the Killing vectors $\xx{0}$ and $\xx{1}$.  Then, the circularity condition implies that $\Gamma$ is orthogonal to $S$. The circularity conditions are met for stationary axisymmetric vacuum and electrovac solutions of the Einstein equations \cite{Kundt:1966zz,Carter:1969zz}. These conditions are also valid for the off-shell metrics (\ref{ds2}), which we shall discuss in the next section.

It is easy to check that
\be
\gamma\equiv -\frac{1}{ 2}\rho_{\mu\nu}\rho^{\mu\nu}=(\lambda_{01})^2-\lambda_{00}\lambda_{11} .
\ee
If one of the Killing vectors vanishes, then the scalar $\gamma$ vanishes as well . For example, this happens in an axisymmetric spacetime at the axis of symmetry. We assume that outside of these points, the Killing vectors are linearly independent, so that they stretch a 2D plane.
If $\gamma>0$, this 2D plane is timelike. For $\gamma<0$, such a 2D plane is spacelike. When $\gamma=0$ and the Killing vectors remain finite, the 2D plane  stretched by these vectors is null.
Carter \cite{Carter:1973rla} showed that if the circularity conditions are satisfied, the event horizon of an arbitrary stationary axially-symmetric black hole coincides with the set of points at which $\rho=0$.

For the case of a stationary axisymmetric black hole in an asymptotically flat spacetime, which is the object of the main interest for us, the region with $\gamma>0$ coincides with the black hole exterior. In this domain there exists a natural metric $\ts{q}$ and alternating tensor $\ts{\epsilon}$ on $S$
\bea
&q_{\mu\nu}=g_{\mu\nu}-\gamma_{\mu\nu}\, ,\\
&\epsilon_{\mu\nu}=\frac{1}{ \sqf} e_{\mu\nu\alpha\beta} \xx{0}{}^{\alpha}\xx{1}{}^{\beta}\, .
\eea
Let $z^a$, $a, b=2,3$ be coordinates in $S$. Then its 2D line element is
\be
dq^2=q_{\mu\nu}dx^{\mu}dx^{\nu}=q_{ab}dz^a dz^b\, .
\ee
Let $p^i$,  $i,j=0,1$ be coordinates in $\Gamma$, then
\be
d\gamma^2=\gamma_{\mu\nu}dx^{\mu}dx^{\nu}=\gamma_{ij}dp^i dp^j\, .
\ee
Denote
\be
\n{DETS}
g=-\mbox{det}(g_{\mu\nu})\hhh \gamma=-\mbox{det}(\gamma_{ij})\hhh q=\mbox{det}(q_{ab})\, ,
\ee
 then one has
 \be \n{ggh}
 g=\gamma q\, .
 \ee
Let us note that $\gamma$ is a scalar function on $S$.

The inverse 2D metric $q^{ab}$ is defined by the relation
\be
q^{ab} q_{bc}=\delta_c^a\, .
\ee
One also has
\be
q_{\mu}^{\nu}=\delta_{\mu}^{\nu}+\frac{1}{ \gamma}\left[
\lambda_{11}\xx{0}_{\mu}\xx{0}{}^{\nu}+\lambda_{00}\xx{1}_{\mu}\xx{1}{}^{\nu}-2\lambda_{01}g^{\nu\kappa}\xx{0}_{(\mu}\xx{1}_{\kappa)}
\right] .
\ee
This tensor is a projector on $S$ which has the following property
\be
q^{\mu}_{\nu}\xx{0}{}^{\nu}= q^{\mu}_{\nu}\xx{1}{}^{\mu}=0\, .
\ee

Let us consider a tensor $\ts{T}$  in $M$ with components $T_{\ldots \alpha\ldots}^{\ldots \beta\ldots} $. We say that this tensor respects the symmetry of $M$ if
\be \n{SSSS}
{\CAL L}_{\inds{\xx{0}}}\ts{T}={\CAL L}_{\inds{\xx{1}}}\ts{T}=0\, .
\ee
Here ${\CAL L}$ is the Lie derivative. We say that a tensor $\ts{T}$ which respects the symmetry is an $S$-tensor if, in addition to (\ref{SSSS}), it has the following property
\be
 \gamma^{\alpha}_{\mu}T_{\ldots \alpha\ldots}^{\ldots \beta\ldots}=0\hh
 \gamma^{\mu}_{\beta}T_{\ldots \alpha\ldots}^{\ldots \beta\ldots}=0\, .
 \ee
Such tensors have non-vanishing components only in the directions tangent to $S$. Geroch \cite{Geroch_2} showed that they can be identified with tensors on $S$, and the covariant derivative $\nabla_{\mu}$ acting on $S$-tensors in $M$ is related to the covariant derivative $D_a$ for the reduced 2D metric $dq^2$ on $S$.

\subsection{Conserved and non-conserved currents}

In what follows we discuss solutions of equation (\ref{MAIN}) for the chiral current in the presence of a chiral anomaly. Let us make comments concerning properties of an arbitrary vector field  $\ts{J}$ in $M$ which respects the spacetime symmetry and therefore obeys the equations
\be \n{LLL}
\CAL{L}_{\inds{\xx{0}}}\ts{J}={\CAL L}_{\inds{\xx{1}}}\ts{J}=0\, .
\ee
Let us write such a vector $\ts{J}$ in the form
\ba\n{EKS}
&J^{\mu}=J_{\ins{K}}^{\mu}+J_\ins{S}^{\mu}\, ,\\
&J_\ins{K}^{\mu}=j_0\xx{0}{}^{\mu}+j_1\xx{1}{}^{\mu}\hh
J_{S}^{\mu}=q^{\mu}_{\ \nu} J^{\nu}\, .
\ea
Since the Killing vectors commute, conditions (\ref{LLL}) imposed on the vector $\ts{J}_\ins{K}$ imply
\be
{\xx{0}}{}^{\mu}\pa_{\mu}j_0={\xx{1}}{}^{\mu}\pa_{\mu}j_0={\xx{0}}{}^{\mu}\pa_{\mu}j_1={\xx{1}}{}^{\mu}\pa_{\mu}j_1=0 .
\ee
This means that the component $J_\ins{K}^{\mu}$ of the vector $J^{\mu}$ is uniquely determined by two scalar functions $j_0$ and $j_1$ on $S$.
It is easy to check that
\be
\nabla_{\mu}(j_0\xx{0}{}^{\mu})=\nabla_{\mu}(j_1\xx{1}{}^{\mu})=0\, .
\ee
This means that the current $\ts{J}_\ins{K}$ satisfies the homogeneous equation
\be
J_{\ins{K}\, ;\mu}^{\mu}=0\, .
\ee

The other component $\ts{J}_\ins{S}$ of the vector $\ts{J}$ is an $S$-vector.
If the vector $\ts{J}$ satisfies the equation (\ref{MAIN}), then $\ts{J}_\ins{S}$ should satisfy this equation as well.
If one finds any special solution of such an inhomogeneous equation, one can obtain a general solution by adding the special solution to solutions of the homogeneous equation
\be \n{Jhom}
J^{\mu}{}_{;\mu}=0\, .
\ee
For the $S$-vector $J^{\mu}=(0,J^r,J^y,0)$,  this equation takes the form
\be \n{JJrryy}
\pa_r(\sqg \,J^r)+\pa_y(\sqg \,J^y)=0\, .
\ee

Let $\ts{J}$ be an $S$-vector obeying the equation (\ref{Jhom}).  Denote
\be
\omega_{\alpha}=e_{\alpha\beta\gamma\delta} J^{\beta}\,\xx{0}{}^{\gamma}\,\xx{1}{}^{\delta}\, .
\ee
Here   $e_{\alpha\beta\gamma\delta}=\sqg\, \epsilon_{\alpha\beta\gamma\delta}$,
and $\epsilon_{\alpha\beta\gamma\delta}$ is the totally antisymmetric Levi-Civita symbol,  $\epsilon_{\tau r y \psi}=1$.
It is easy to check that $\ts{\omega}$ is an $S$-covector and its components are
\be\nonumber
\omega_{\mu}=(0,\omega_r,\omega_y,0)\, ,\  \omega_r=\sqg \,J^y\, ,\  \omega_y=-\sqg \,J^r\, .
\ee
If the $S$-current satisfies the conservation law (\ref{JJrryy}), then one has
\be
\omega_{r ,y}=\omega_{y ,r}\, .
\ee
This means that for a conserved $S$-current $\ts{J}$, the 1-form $\ts{\omega}$ is closed and (at least locally) is a gradient of some function $\Psi=\Psi(r,y)$
\be
\omega_{\mu}=\Psi_{,\mu}\, .
\ee
We call $\Psi$ the potential. The conserved $S$-current itself can be expressed in terms of its potential as follows
\ba\n{JPSI}
J^{\mu}=-\frac{1}{\gamma}e^{\mu\nu\rho\sigma}\Psi_{,\nu}\,\xx{0}{}_{\rho}\,\xx{1}{}_{\sigma}\, .
\ea

\section{Hidden symmetry and canonical form of the metric}
\n{s3}
\subsection{Principal Killing-Yano tensor and off-shell metric}

Consider a non-degenerate rank-2 skew-symmetric tensor $h_{\mu\nu}$ obeying the equation
\be \n{hhh}
h_{\mu\nu ;\lambda}=g_{\lambda\nu}\xi_{\mu}-g_{\lambda\mu}\xi_{\nu}\, .
\ee
It is a closed conformal Killing-Yano tensor and it is called a principal Killing-Yano tensor. The 2-form $\ts{h}$ is closed and (at least locally) it can be written as
\be
\ts{h}=d\ts{b}\, ,
\ee
where $\ts{b}$ is a 1-form. If $\ts{h}$ is a principal Killing-Yano tensor, then
\begin{enumerate}[label=(\roman*)]
\item The Hodge-dual of $\ts{h}$, $\ts{k}=\star\ts{h}$, is a Killing-Yano tensor
\be
k_{\mu (\nu ;\lambda)}=0\, .
\ee
\item The following two symmetric tensors $K_{\mu\nu}=k_{\mu\lambda}k^{\lambda}_{\ \nu}$
    and $H_{\mu\nu}=h_{\mu\lambda}h^{\lambda}_{\ \nu}$ are Killing and conformal Killing tensors, respectively
    \be
    K_{(\mu\nu ;\lambda)}=0
    \hh
      H_{(\mu\nu ;\lambda)}=2g _{(\mu\nu}h_{\lambda)\sigma}\xi^{\sigma} .
    \ee
\end{enumerate}

In the general case, the metric of a spacetime admitting the principal Killing-Yano tensor can be written in the following form
(see e.g. \cite{Frolov:2017kze,FrolovZelnikov:2011})
\ba\label{ds2}
&ds^2=d\gamma^2+dq^2\, ,\\
&d\gamma^2=\gamma_{ij}dp^i dp^j=\frac{\Delta_y}{\Sigma}
(d\tau-r^2d\psi)^2-\frac{\Delta_r}{\Sigma}(d\tau+y^2d\psi)^2\, .\\
&dq^2=q_{ab}dz^a dz^b=\Sigma\Big[\frac{dr^2}{\Delta_r}+\frac{dy^2}{\Delta_y}\Big] .
\ea
Here $p^i=(\tau,\psi)$, $z^a=(r,y)$, $\Sigma=r^2+y^2$, $\Delta_r=\Delta_r(r)$ and $\Delta_y=\Delta_y(y)$.
This metric is called ``off-shell" if the functions $\Delta_r$ and $\Delta_y$ are not specified.
For this metric, one has
\be\nonumber
\sqg=\Sigma\,, ~~~ \sqf=\sqrt{\Delta_r\Delta_y}\,, ~~~   \sqq=\frac{\Sigma}{\sqrt{\Delta_r\Delta_y}} \,,
\ee
using the conventions adopted in \eqref{DETS}. In the coordinates $(\tau,r,y,\psi)$, the 1-form $\ts{b}$ is of the form
\be
\ts{b}=\frac{1}{2}\left[(r^2-y^2)d\tau+r^2y^2d\psi\right]\, .
\ee

The metric (\ref{ds2}) has two commuting Killing vectors $\ts{\xx{0}}=\pa_{\tau}$ and $\ts{\xx{1}}=\pa_{\psi}$. The first one $\ts{\xx{0}}$ coincides with the vector $\ts{\xi}$ which enters the definition (\ref{hhh}) of the principal Killing-Yano tensor, while the second Killing vector $\ts{\xx{1}}$ satisfies the equation $\xx{1}_{\mu}=K_{\mu\nu}{\xx{0}}{}^{\nu}$. The metric (\ref{ds2}) is invariant under a discrete transformation $\tau\to -\tau$, $\psi\to -\psi$.

Consider the following equation
\be \n{EIG}
H^{\mu}_{\ \nu}u^{\nu}=\lambda u^{\mu}\,.
\ee
The eigenvalues $\lambda$ are $r^2$ and $-y^2$. These parameters $(r,y)$ together with Killing parameters $(\tau,\psi)$ enter in the metric (\ref{ds2}) as its coordinates. The full metric (\ref{ds2}) evidently belongs to the class of the metrics discussed in \cite{Geroch_2}. The Darboux coordinates $(r,y)$ define standard coordinates in the 2D space $S$.

For the metric (\ref{ds2}) there exists a so-called ``Darboux basis" of normalized vectors $\ts{e}_A=\{ \ts{e}_{\bi{1}}, \ts{e}_{\bbi{1}}, \ts{e}_\bi{2},\ts{e}_{\bbi{2}}\}$ in which
\ba
&g_{\mu\nu}=\eta^{AB}e_{A\mu}e_{B\nu}\, ,\quad \eta^{AB}=\mbox{diag} (-1,1,1,1)\, ,\\
&h_{\mu\nu}=2\left( r e_{\bi{1}[\mu} {e}_{\bbi{1} \nu]}+y e_{\bi{2}[\mu}{e}_{\bbi{2} \nu]}\right)\, .
\ea
We use the capital letters such as $A$ and  $B$ to enumerate the Darboux basis vectors. These indices take the values $(1,\bar{1},2,\bar{2})$.

The vectors of the Darboux basis are
\ba\n{DDEE}
&e_{\bi{1}}^{\mu}\pa_{\mu}=\frac{1}{\Sigma}\Big(\frac{ \Sigma}{ \Delta_r}\Big)^{1/2}
(r^2 \pa_{\tau}+\pa_{\psi} )  ,    \\
& e_{\bbi{1}}^{\mu}\pa_{\mu}=\Big(\frac{ \Delta_r}{\Sigma }\Big)^{1/2} \pa_r  , \\
&e_{\bi{2}}^{\mu}\pa_{\mu}=\frac{1}{\Sigma}\Big(\frac{\Sigma}{\Delta_y}\Big)^{1/2}(y^2\pa_{\tau}-\pa_{\psi}) ,\\
&e_{\bbi{2}}^{\mu}\pa_{\mu}=\Big(\frac{\Delta_y}{ \Sigma}\Big)^{1/2}\pa_y  \, ,
\ea
\ba\n{DDFF}
&e_{\bi{1}\mu}dx^{\mu}=-\Big(\frac{\Delta_r}{\Sigma}\Big)^{1/2}(d\tau +y^2 d\psi) ,\\
& e_{\bbi{1}\mu}dx^{\mu}=\Big(\frac{ \Sigma}{\Delta_r}\Big)^{1/2} dr , \\
&e_{\bi{2}\mu}dx^{\mu}=\Big(\frac{\Delta_y}{\Sigma}\Big)^{1/2}(d\tau -r^2 d\psi) ,\\
&e_{\bbi{2}\mu}dx^{\mu}=\Big(\frac{\Sigma}{\Delta_y}\Big)^{1/2}dy \, .
\ea
Here we choose $e_{\bi{1}}^{\mu}$ to be a future-directed timelike vector, and we choose the orientation of the tetrad such that
\ba
e_{\alpha\beta\gamma\delta} \,  e_{\bi{1}}^{\alpha} e_{\bbi{1}}^{\beta} e_{\bi{2}}^{\gamma} e_{\bbi{2}}^{\delta}= +1 .
\ea
It is easy to show that
\ba
&{e}_{\bi{1}}^{\mu}=\frac{1}{\sqrt{|\lambda_{1}|}} h^{\mu}_{\ \nu}{e}_{\bbi{1}}^{\nu}\hh {e}_{\bi{2}}^{\mu}=\frac{1}{\sqrt{|\lambda_{2}|}} h^{\mu}_{\ \nu}{e}_{\bbi{2}}^{\nu}\, ,
\ea
where $\lambda_{1}=r^2$ and $\lambda_{2}=-y^2$ are the eigenvalues of equation \eqref{EIG}.

\begin{figure}[!htb]%
    \centering
    \includegraphics[width=0.35\textwidth]{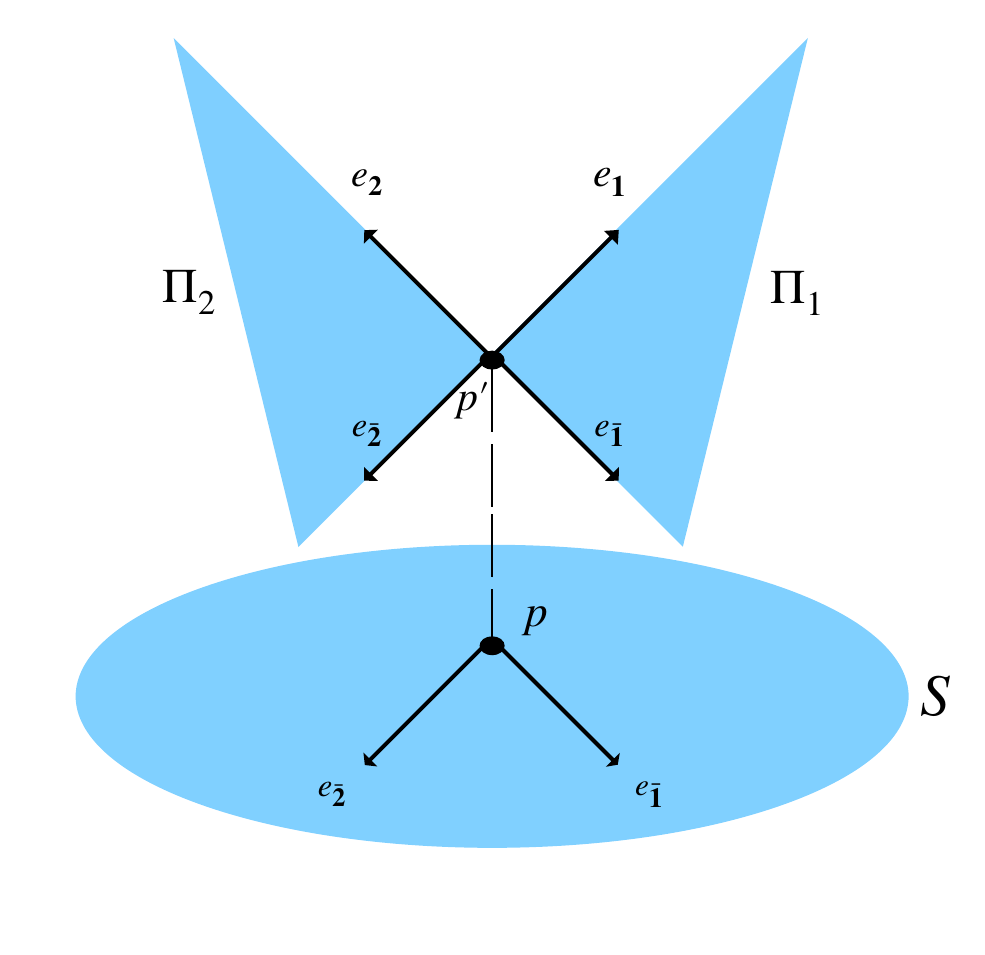}\\[-30pt]
    \caption{Illustration of the geometry of spacetime for the off-shell metric (\ref{ds2}). }
    \label{Fig_1}
\end{figure}

The principal Killing-Yano tensor and the vectors of the Darboux basis obey the following relations
\ba
&{\CAL L}_{\inds{\xx{0}}}\ts{h}={\CAL L}_{\inds{\xx{1}}}\ts{h}=0\, ,\\
&{\CAL L}_{\inds{\xx{0}}}\ts{e}_A={\CAL L}_{\inds{\xx{1}}}\ts{e}_A=0\, .
\ea
The vectors $\ts{e}_{\bbi{1}}$ and $\ts{e}_{\bbi{2}}$ lie in $S$, while the vectors $\ts{e}_{\bi{1}}$ and $\ts{e}_{\bi{2}}$ are linear combinations of the Killing vectors $\xx{0}$ and $\xx{1}$ with coefficients depending on $r$ and $y$.
We denote by $\Pi_1$ a two-plane spanned by the vectors $\ts{e}_{\bi{1}}$ and $ \ts{e}_{\bbi{1}}$ and by $\Pi_2$ a two-plane spanned by the vectors $\ts{e}_{\bi{2}}$ and $ \ts{e}_{\bbi{2}}$. (See figure~\ref{Fig_1}).
Vectors of $\Pi_1$  are eigenvectors of the tensors $\ts{H}$ and $\ts{K}$ with eigenvalues  $\lambda=r^2$ and  $\lambda=y^2$, respectively. Similarly, vectors of $\Pi_2$  are eigenvectors of the tensors $\ts{H}$ and $\ts{K}$ with eigenvalues  $\lambda=-y^2$ and  $\lambda=-r^2$, respectively.

Using the Darboux  basis vectors, we define
\ba\n{kkll}
\ts{k}_{\pm}=\ts{e}_{\bi{1}} \mp \ts{e}_{\bbi{1}}\hh \ts{l}_{\pm}=\sqrt{\frac{ \Sigma}{ \Delta_r}}\ts{k}_{\pm}
\, ,
\ea
\ba
l^{\mu}_{\pm}\partial_{\mu}=\frac{r^2}{\Delta_r}\partial_{\tau}+\frac{1}{\Delta_r}\partial_{\psi}   \mp   \partial_r .
\ea
The vectors $\ts{k}_{\pm}$ and  $\ts{l}_{\pm}$ are null. The vectors $\ts{k}_{\pm}$  obey the following normalization condition
\be
(\ts{k}_+,\ts{k}_-)=-2\, .
\ee
The vectors $\ts{l}_{\pm}$ satisfy the relation
\be
l^{\nu}_{\pm} l^{\mu}_{\pm ;\nu}=0\, .
\ee
This relation shows that the integral lines of $\ts{l}_{\pm}$ are geodesics, and that $r$ is an affine parameter along them.
These vectors are called principal null vectors.

\subsection{Black hole metrics}

Functions $\Delta_r(r)$ and $\Delta_y(y)$ which enter the off-shell metric (\ref{ds2}) are arbitrary. They are specified if one requires that this metric is a solution of the Einstein equations. For the vacuum, such a solution is the Kerr metric, while for the electrovacuum it is the Kerr-Newman metric. It is instructive to work initially with a general form of the metric and specify the functions $\Delta_r(r)$ and $\Delta_y(y)$ later. However, we assume that the function $\Delta_r$ obeys special boundary conditions
\begin{enumerate}[label=(\roman*)]
\item $\Delta_r$ is positive in the interval $r_H<r<\infty$.
\item At $r\to \infty$ $\Delta_r$ has the following asymptotic form $\Delta_r\sim r^2-2mr+\ldots$.
\item The metric function $\Delta_r(r)$ vanishes at $r=r_H$, and near this point it has the following expansion $\Delta_r\sim \Delta'_r (r_H)(r-r_H)+O((r-r_H)^2)$.
\end{enumerate}

In order to reproduce the standard formulas in the Boyer-Lindquist coordinates, we introduce two new Killing coordinates $(t,\phi)$
\be
\tau=t-a\phi\hh \psi=\phi/a\, .
\ee
We denote by $\ts{\xi}_{(t)}=\pa_t$ and $\ts{\xi}_{(\phi)}=\pa_{\phi}$ their corresponding Killing vectors. Then, one has
\be
\ts{\xi}_{(t)}=\ts{\xx{0}}=\pa_{\tau}\hh \ts{\xi}_{(\phi)}=\frac{1}{ a}\ts{\xx{1}}-a\ts{\xx{0}}\ .
\ee
The Killing vector $\ts{\xi}_{(t)}$ is singled out by the property that its norm at infinity is finite and  equal to $-1$. The other Killing vector $\ts{\xi}_{(\phi)}$ has the property that its integral lines are closed. The axes of symmetry are defined by the condition $\ts{\xi}_{(\phi)}^2=0$. This condition implies that $\Delta_y$ vanishes at the symmetry axis.

For
\be \n{DY}
\Delta_y=a^2-y^2
\ee
 the asymptotic form of the metric at infinity is
\be
ds^2\approx -dt^2+dr^2+r^2(d\theta^2+\sin^2\theta d\phi^2)+\dots  .
\ee
A condition for regularity at the symmetry axes (absence of conical singularities) implies that $\phi$ is an angle variable and its period is $2\pi$.

In what follows we assume that relation (\ref{DY}), as well as the imposed conditions on $\Delta_r$, are satisfied.
Such a metric describes a rotating black hole in an asymptotically flat spacetime.
Let us emphasize that in the presence of matter in the black hole exterior, the metric function $\Delta_r$ should be obtained by solving the Einstein equations.

The angular velocity of a black hole described by the metric (\ref{ds2}) can be defined as follows. Denote
\be \n{EETA}
\ts{\eta}=\ts{\xi}_{(t)}+\Omega \ts{\xi}_{(\phi)} =(1-a\Omega)\ts{\xx{0}}+\frac{\Omega}{a}\ts{\xx{1}} .
\ee
In the black hole exterior, the condition $\ts{\eta}^2=0$ specifies two values of $\Omega$. The two corresponding null vectors coincide at the horizon. In this limit
\be
\Omega=\Omega_H=\frac{a}{ r_H^2+a^2}\, .
\ee
 Here, $\Omega_H$ is the angular velocity of the black hole.
 Inserting this value of $\Omega$ into the definition of $\ts{\eta}$, one can check that the following relation is valid at the horizon
 \be\n{surf}
 (\ts{\eta}^2)_{,\mu}=-2\kappa \eta_{\mu} ,
 \ee
 where $\kappa$ is the surface gravity. The validity of this vector equation should be checked in coordinates which are regular at the horizon.
One can also compute $\kappa$ using the following representation
 \ba
 \kappa^2=-\frac{1}{2}\eta^{\alpha;\beta}\eta_{\alpha;\beta}\big|_H \, .
 \ea
This gives the following expression for the surface gravity
\be\label{kappa1}
\kappa=\frac{\Delta_r'\big|_H}{ 2(r_H^2+a^2)} \, .
\ee
Here prime denotes the derivative with respect to $r$.

\subsection{Vacuum and electrovac metrics}

For a given matter distribution, the Einstein equations impose restrictions on the functions $\Delta_r(r)$ and $\Delta_y(y)$. For example, let us assume that the scalar curvature vanishes. Then, one has
\be
R=-\frac{1}{\Sigma}\left( \frac{d^2\Delta_r}{ dr^2}+ \frac{d^2\Delta_y}{ dy^2}\right)=0\, .
\ee
This relation implies that both functions $\Delta_r$ and $\Delta_y$ are quadratic polynomials of their arguments. A term in $\Delta_y$ that is linear in $y$ is connected with the NUT parameter. In the presence of this parameter, the metric has  at least one conical
singularity at the poles of the axis of rotation (north or south).
We assume that this singularity is absent, and write
\be
\Delta_y=a^2-y^2\, .
\ee
This is the form of the metric function $\Delta_y$ that we postulated earlier.

We assume that the quadratic equation $\Delta_r=0$ has two positive roots $r_-<r_+$. It is convenient to write  $\Delta_r$ in the form
\be\n{KKNN}
\Delta_r=(r-m)^2-b^2\hh r_{\pm}=m\pm b\, .
\ee
The location of the black hole horizon coincides with $r_+$, thus one has $r_H=r_+$.
For the vacuum solution, $b=\sqrt{m^2-a^2}$, and one obtains the Kerr metric. For the electrovac metric, the scalar curvature vanishes and the expression (\ref{KKNN}) is valid where
\be
b=\sqrt{m^2-a^2-Q^2-P^2}\, .
\ee
This is the metric of a charged rotating black hole in an asymptotically flat spacetime. The parameters $m$, $ma$, $Q$ and $P$ are its mass, angular momentum, electric monopole charge, and magnetic monopole charge, respectively. For $Q=P=0$ one reproduces the Kerr metric.

\subsection{Regularity conditions}\label{Regularity}

\subsubsection{Coordinates regular at the horizon}

The metric (\ref{ds2}) is singular at the points where either $\Delta_r$ or $\Delta_y$ vanishes. The surface where $\Delta_r=0$ is a Killing horizon, while the condition $\Delta_y=0$ defines an axis of rotation. In what follows, we assume that the surface gravity $\kappa$ defined by (\ref{surf}) is finite, so that the horizon is non-degenerate.
In this case, the singularity of the metric at $\Delta_r=0$ is a coordinate singularity. In order to obtain coordinates which are regular either at the future or the past horizon, one can introduce coordinates similar to Kerr's Eddington-Finkelstein retarded and advanced time coordinates. For this purpose, let us denote
\ba
& d\sigma_{\pm}=d\tau+a^2 d\psi \pm \frac{r^2+a^2}{\Delta_r} dr ,\\
&d\phi_{\pm}=a\Big(d\psi\pm \frac{1}{\Delta_r} dr\Big).
\ea
Then one has
\ba
& d\tau=d\sigma_{\pm}-a d\phi_{\pm} \mp \frac{r^2}{\Delta_r} dr\, ,\\
&d\psi =\frac{1}{a}d\phi_{\pm} \mp \frac{1}{\Delta_r}dr  .
\ea
In the vicinity of the horizon, one has
\ba
\Delta_r&= \frac{d\Delta_r}{ dr}\Big|_{r=r_+}(r-r_+)+O( (r-r_+)^2)\nonumber \\
&=2\kappa (r_H^2+a^2) (r-r_+)+O( (r-r_+)^2) ,
\ea
where $\kappa$ is the horizon surface gravity.

In $(\sigma_{\pm},\phi_{\pm})$ coordinates, the metric (\ref{ds2}) near the horizon takes the following form
\ba\n{regm}
 ds^2&=\frac{\Delta_y}{\Sigma}\Big( d\sigma_{\pm}
 -\frac{r^2+a^2}{a} d\phi_{\pm}\Big)^2\\
 &\pm 2 \Big( d\sigma_{\pm}-\frac{a^2-y^2}{a} d\phi_{\pm}\Big)dr+\frac{\Sigma}{\Delta_y} dy^2+\ldots \, .
\ea
Here $(\ldots)$ denotes terms which vanish on the horizon. This expression demonstrates the regularity of the metric at the horizon in these new coordinates $\sigma_{\pm}$ and $\phi_{\pm}$.
Namely, the advanced time coordinates $\sigma_+$ and $\phi_+$ are regular at the future horizon $H_+$, while $\sigma_-$ and $\phi_-$ are regular at the past horizon $H_-$.

\subsubsection{Regular vectors}

\begin{figure}[!htb]%
    \centering
    \includegraphics[width=0.25\textwidth]{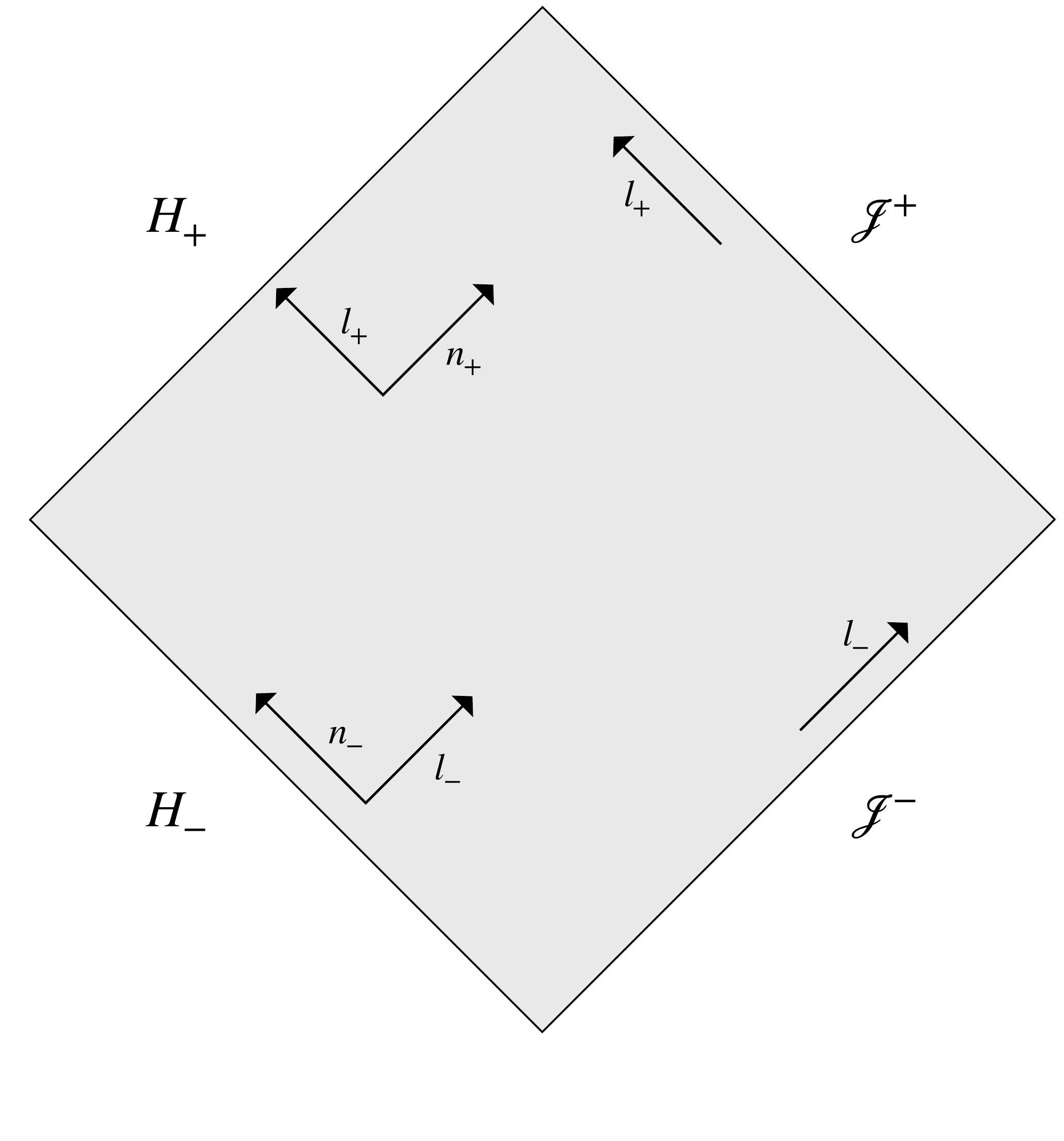}\\[-20pt]
    \caption{Vectors $\ts{l}_{\pm}$ and $\ts{n}_{\pm}$.}
    \label{Fig_2}
\end{figure}

A tensor is regular at the horizon $H_{\pm}$ if its components in $(\sigma_{\pm},\phi_{\pm})$ coordinates are finite and smooth at $H_{\pm}$.
It is easy to show that in the regular coordinates $(\sigma_{\pm},\phi_{\pm})$ the components of $\ts{l}_{\pm}$ are
\be
l_{\pm \mu}dx^{\mu}=-d\sigma_{\pm}+\frac{a^2-y^2}{ a} d\phi_{\pm}\, .
\ee
This means that the principal null vector $\ts{l}_+$ is regular at $H_+$, while  $\ts{l}_-$ is regular at $H_-$. Let us denote
\be
\ts{n}_{\pm}=\sqrt{\frac{\Delta_r}{ \Sigma}} \ts{k}_{\mp}\, .
\ee
One has
\be
(\ts{l}_{\pm},\ts{n}_{\pm})=-2\, .
\ee
The null vectors $\ts{n}_+$ and $\ts{n}_-$ are regular at the future $H_+$ and at the past $H_-$ horizons, respectively. Their  components in $(\sigma_{\pm},\phi_{\pm})$ coordinates are
\be
n_{\pm \mu}dx^{\mu}= -\frac{\Delta_r}{\Sigma}(d\sigma_{\pm}-\frac{a^2-y^2}{a} d\phi_{\pm})  \pm 2 dr\, .
\ee
At the horizons $H_{\pm}$, the vectors $\ts{n}_{\pm}$ coincide with the  null generators of the horizon.
Figure~\ref{Fig_2} shows vectors $\ts{l}_{\pm}$ and $\ts{n}_{\pm}$.

\subsubsection{Regularity at the symmetry axes}

Consider an $S$-vector $\ts{J}$, and let $J^y$ be its $y$-component.
Let us denote
\be\n{YYY}
Z=\frac{J^y}{ \Delta_y}\, .
\ee
A condition for the vector current $\ts{J}$ to be regular at the axis of symmetry is that the limit of the ratio on the right-hand side of (\ref{YYY}) at $y=\pm a$ exists, and that the function $Z$ is well defined. This condition follows from the requirement that for fixed $t$ and $r$, a 2D section is a locally-flat 2-plane, and $\ts{J}$ is regular in its 2D Cartesian coordinates.


\section{Solving the chiral anomaly equation}
\n{s4}
\subsection{Chiral currents respecting explicit and hidden symmetries}

We now study solutions of the chiral anomaly equation (\ref{MAIN}) with the right-hand side equal to the Pontryagin pseudoscalar $ \mathcal{P}=-\frac{1}{2}R_{\mu\nu\alpha\beta}{}^*R^{\mu\nu\alpha\beta}$.  We consider first a general case where the spacetime geometry is described by an off-shell metric (\ref{ds2}), and specify these solutions for particular black hole metrics later.

Calculations give the following expression for the Pontryagin invariant in the off-shell metric (\ref{ds2})
\ba\n{PONT}
\mathcal{P}=&\frac{2}{\Sigma^6}\Big\{\big[
\Sigma (r\,\dot{\Delta}_y-y\,\Delta_r')+4ry(\Delta_{r}-\Delta_{y})
\big]\\
&\times
\big[
-\Sigma^2 (\Delta_r''+\ddot{\Delta}_y)
+6\Sigma (r\,\Delta_r'+y\,\dot{\Delta}_y)\\
&-12(r^2-y^2)(\Delta_{r}-\Delta_{y})
\big]
\Big\} .
\ea
Here and later on we use a prime and a dot to denote the derivatives with respect to $r$ and $y$, respectively.

First we focus on the solutions of the chiral anomaly equation which respect the spacetime symmetry. As it was demonstrated in section~\ref{s2}, such current vectors can be written in the form (\ref{EKS})
\be
J^{\mu}=J_K^{\mu}+J_S^{\mu}\, .
\ee
The chiral current $J_{K}^{\mu}$ is a solution of the homogeneous equation, and has nonvanishing components only in $(\tau,\psi)$ directions. The other current $J_S^{\mu}$ is an $S$-vector, and it has components $J_S^{\mu}=(0,J^r,J^y,0)$, where $J^r$ and $J^y$ are functions of $(r,y)$. Hence, the equation (\ref{MAIN}) takes the form
\be\n{SMAIN}
\pa_r(\Sigma J_S^r)+\pa_y(\Sigma J_S^y)=\Sigma \mathcal{P}\, .
\ee

Let us assume that the $S$-vector of the chiral current $\ts{J}_S$ satisfies an additional property. Namely, it is an eigenvector of the Killing tensor $K_{\mu\nu}$
\be \n{KKKJJ}
K^{\mu}_{\ \nu}J_S^{\nu}=\lambda J_S^{\mu}\, .
\ee
This condition implies that either the $J^r$ or $J^y$ component of the current vanishes.
There exist two linearly independent $S$-currents which obey this property. We denote them by $\ts{J}_{(r)}$ and $\ts{J}_{(y)}$. The eigenvector $\ts{J}_{(r)}$ has components  $(0,J^r,0,0)$, and the corresponding eigenvalue is $\lambda=y^2$. The other eigenvector $\ts{J}_{(y)}$ has components  $(0,0,J^y,0)$, and the corresponding eigenvalue is $\lambda=-r^2$.
We say that the current respects the hidden symmetry if it satisfies condition (\ref{KKKJJ}).

To distinguish between the currents $\ts{J}_{(r)}$ and $\ts{J}_{(y)}$, we call them $R$-current and $Y$-current, respectively.
Both currents are $S$-vectors and, hence, they are spacelike. This implies that in the Darboux reference frame, their temporal components vanish, and hence, the net chiral charge density vanishes as well.

The currents $\ts{J}_{(r)}$ and $\ts{J}_{(y)}$ can be found by integration of the following equations
\be
\pa_r(\Sigma J_{(r)}^r)=\Sigma \mathcal{P}  \hh \pa_y(\Sigma J_{(y)}^y)=\Sigma \mathcal{P}\, .
\ee
We write these solutions in the form
\ba\n{JJrrr}
&J^{\mu}_{(r)}=\frac{R-R_0(y)}{ \Sigma} \delta^{\mu}_r\hh
R=\int dr \Sigma \mathcal{P}\, ,\\
&J^{\mu}_{(y)}=\frac{Y-Y_0(r)}{ \Sigma} \delta^{\mu}_y\hh
Y=\int dy \Sigma \mathcal{P}\, .
\ea

Calculations give
\ba\n{RRYY}
R(r,y)&=\frac{1}{\Sigma^4}\Big\{
y\Sigma^2(\Delta_r')^2-3y\Sigma^2(\dot{\Delta}_y)^2\\
&-2r\Sigma^2 \Delta_r' \dot{\Delta}_y-8ry\Sigma(\Delta_r-\Delta_y)\Delta_r'\\
&+(6\Sigma-16y^2)\Sigma(\Delta_r-\Delta_y)\dot{\Delta}_y \\
&+\Sigma^2[2y(\Delta_r-\Delta_y)+\Sigma\dot{\Delta}_y]\ddot{\Delta}_y\\
&+8y(\Delta_r-\Delta_y)^2(2\Sigma-3y^2)
\Big\}\, ,\\
Y(r,y)&=\frac{1}{\Sigma^4}\Big\{
-r\Sigma^2(\dot{\Delta}_y)^2+3r\Sigma^2(\Delta_r')^2\\
&+2y\Sigma^2 \Delta_r' \dot{\Delta}_y-8ry\Sigma(\Delta_r-\Delta_y)\dot{\Delta}_y\\
&-(10\Sigma-16y^2)\Sigma(\Delta_r-\Delta_y)\Delta_r' \\
&+\Sigma^2[2r(\Delta_r-\Delta_y)-\Sigma\Delta_r']\Delta_r''\\
&+8r(\Delta_r-\Delta_y)^2(\Sigma-3y^2)
\Big\}\, ,
\ea

It is easy to check that for $\Delta_y=a^2-y^2$, the functions $R$ and $Y$ have the following properties
\be \n{SYMRY}
R(r,-y)=-R(r,y)\hh Y(r,-y)=Y(r,y)\, .
\ee

The solutions (\ref{JJrrr}) contain two arbitrary functions of one variable, $R_0(y)$ and $Y_0(r)$, which arise as the corresponding ``integration constants". It is easy to check that any current of the form
\be \n{zero}
J_0^{\mu}=(0,R_0(y)/\Sigma, Y_0(r)/\Sigma,0)
\ee
is a solution of the homogeneous equation $J^{\mu}_{0 ;\mu}=0$. We call such solutions ``zero modes".
In section~\ref{s2}, it was shown that  a conserved $S$-current has a potential $\Psi$, and can be written in the form (\ref{JPSI}). The corresponding potential for zero modes (\ref{zero}) is
\be
\Psi=\int Y_0(r)dr-\int R_0(y)dy\, .
\ee

Let us note that functions $R$ and $Y$ which enter solutions (\ref{RRYY}) are finite both at the horizon, where $\Delta_r=0$, and the symmetry axes, where $\Delta_y=0$. However, this does not guarantee that the corresponding solutions $J^{\mu}_{(r)}$ and $J^{\mu}_{(y)}$ are regular. Zero modes determined by functions $R_0(r)$ and $Y_0(r)$ can be used to ``improve" the properties of the solutions (\ref{RRYY}) and to make them regular both at the horizon and at the symmetry axes.

Let $r=r_H$ be a solution of the equation $\Delta_r(r_H)=0$, and let $y=\pm a$ be a solution of the equation $\Delta_y(y=\pm a)=0$.
Let us choose
\be
R_0=R_H\equiv R(r=r_H,y)\, ,\  Y_0=Y_A\equiv Y(r,y=\pm a) .
\ee
Then the current $\ts{J}_{(r)}$ is regular at the horizon where $r=r_H$, and the current $\ts{J}_{(y)}$ is regular at the axes of the rotation where $y=\pm a$. Let us note that $Y(r,-y)=Y(r,y)$, so it is sufficient to make the current $\ts{J}_{(y)}$ regular at one of the axes, say $y=-a$, and it will automatically be regular at the other, $y=a$.

Using expressions for functions $R(r,y)$ and $Y(r,y)$ for the currents  (\ref{RRYY}) one finds
\ba\n{AAA}
R_H&=\frac{1}{\Sigma^4}\Big\{
y\Sigma^2(\Delta_r')^2-3y\Sigma^2(\dot{\Delta}_y)^2\\
&-2r\Sigma^2 \Delta_r' \dot{\Delta}_y+8ry\Sigma\Delta_y\Delta_r'\\
&-(6\Sigma-16y^2)\Sigma\Delta_y\dot{\Delta}_y \\
&+\Sigma^2[-2y\Delta_y+\Sigma\dot{\Delta}_y]\ddot{\Delta}_y\\
&+8y(\Delta_y)^2(2\Sigma-3y^2)
\Big\}\Big|_{r=r_H}\, ,\\
Y_A&=\frac{1}{\Sigma^4}\Big\{
-r\Sigma^2(\dot{\Delta}_y)^2+3r\Sigma^2(\Delta_r')^2\\
&+2y\Sigma^2 \Delta_r' \dot{\Delta}_y-8ry\Sigma\Delta_r\dot{\Delta}_y\\
&+(10\Sigma-16y^2)\Sigma\Delta_r\Delta_r' \\
&-\Sigma^2[2r\Delta_r+\Sigma\Delta_r']\Delta_r''\\
&+8r\Delta_r^2(\Sigma-3y^2)
\Big\}\Big|_{y=a}\, ,
\ea

\subsection{Principal chiral current}

\n{ssPCC}

Using relation (\ref{JJrrr}), one can write
\be
\ts{J}_{(r)}=J_{\bbi{1}} \ts{e}_{\bbi{1}}\, .
\ee
It is easy to check that for an arbitrary function $f=f(r,y)$, one has
\be
(fe_{\bi{1}}^{\mu})_{;\mu}=0\, ,
\ee
Using this property one can modify the current $\ts{J}_{(r)}$ by adding a term proportional to $\ts{e}_{\bi{1}}$.
Such a new current is not an $S$-vector, however it belongs to the $\Pi_1$  plane, and hence it is still an eigenvector of the Killing tensor $\ts{K}$ (see figure~\ref{Fig_1}). The  plane $\Pi_1$ contains two principal null vectors $\ts{l}_{\pm}$, (\ref{kkll}), and by proper choice of coefficients in the linear combinations of  $\ts{J}_{(r)}$ and $\ts{e}_{\bi{1}}$, one can construct chiral currents that are parallel to the principal null vectors.
 For this purpose, let us note that $\ts{e}_{\bbi{1}}\pm \ts{e}_{\bi{1}}$ are null vectors. We define
\be
\ts{J}_{\pm}=\mp \frac{R-R_{\pm}(y)}{ \Sigma} \ts{l}_{\pm}\, .
\ee
Here $R(r,y)$ is a function defined by (\ref{RRYY}), and $R_{\pm}(y)$ are arbitrary functions of $y$.
We call $\ts{J}_{\pm}$ principal chiral currents. These currents are solutions of the chiral anomaly equation (\ref{MAIN}) that respect explicit and hidden spacetime symmetries.

At the infinity $r\to\infty$, $R(r,y)$  vanishes quite fast while $\Sigma\approx r^2$, and so one has
\be\n{ASSJ}
\ts{J}_{\pm}\approx \pm \frac{R_{\pm}(y)}{ r^2}\ts{l}_{\pm}\, .
\ee
Let us consider this chirality current far away from the black hole, in the asymptotically flat domain where
\be
\ts{l}_{\pm}=\pa_t \mp \pa_r\, .
\ee
Let us write the current $\ts{J}_{\pm}$ in the form $J^{\mu}_{\pm}=(-\rho_{\pm},\vec{j}_{\pm})$, where
\be
\rho_{\pm}=-J_{\pm \mu}\xi_{(t)}^{\mu}
\ee
is the chirality density, which for the current (\ref{ASSJ}) is
\be
\rho_{\pm}=\pm \frac{R_{\pm}}{ r^2}\, .
\ee
The spatial components of the chiral currents (\ref{ASSJ}) are
\be
\vec{j}_{\pm}=-\frac{R_{\pm}}{ r^2}\pa_r=\mp \rho_{\pm}\pa_r \, .
\ee
Hence, for a principal chiral current $\ts{J}_+$, the function $R_{+}(y)$ describes the intensity of the incoming chirality flux at ${\cal J}^-$, while $-R_-(y)$ is the intensity of the outgoing chirality flux at ${\cal J}^+$. When $R_{\pm}=0$  the corresponding fluxes vanish.

This property illustrates a main difference between the principal chiral currents and the $R$- and $Y$-currents described above. Namely, the principal currents can describe fluxes of the chirality at infinity.
For example, for Hawking radiation there exists a spatial separation of created particles with different chirality (see e.g.  \cite{Leahy:1979,Vilenkin:1979,Bolashenko:1989,Bolashenko:1989trudy,Casals_2009}).
Let us note that the principal current $\ts{J}_{+}$ vanishes at the future horizon $H_+$ when $R_{+}=R_H$, and the principal current $\ts{J}_{-}$ vanishes at the past horizon $H_-$ when $R_{-}=R_H$.

As we already mentioned, there exists a wide ambiguity in the choice of a solution to the chiral anomaly equation (\ref{MAIN}).
This reflects an ambiguity in the choice of the system's state. The choice of a special state imposes restrictions on the initial and/or boundary conditions for the chiral currents. Let us consider 3 important cases.

\subsubsection{$B$-state}

\begin{figure}[!htb]%
    \centering
    \includegraphics[width=0.25\textwidth]{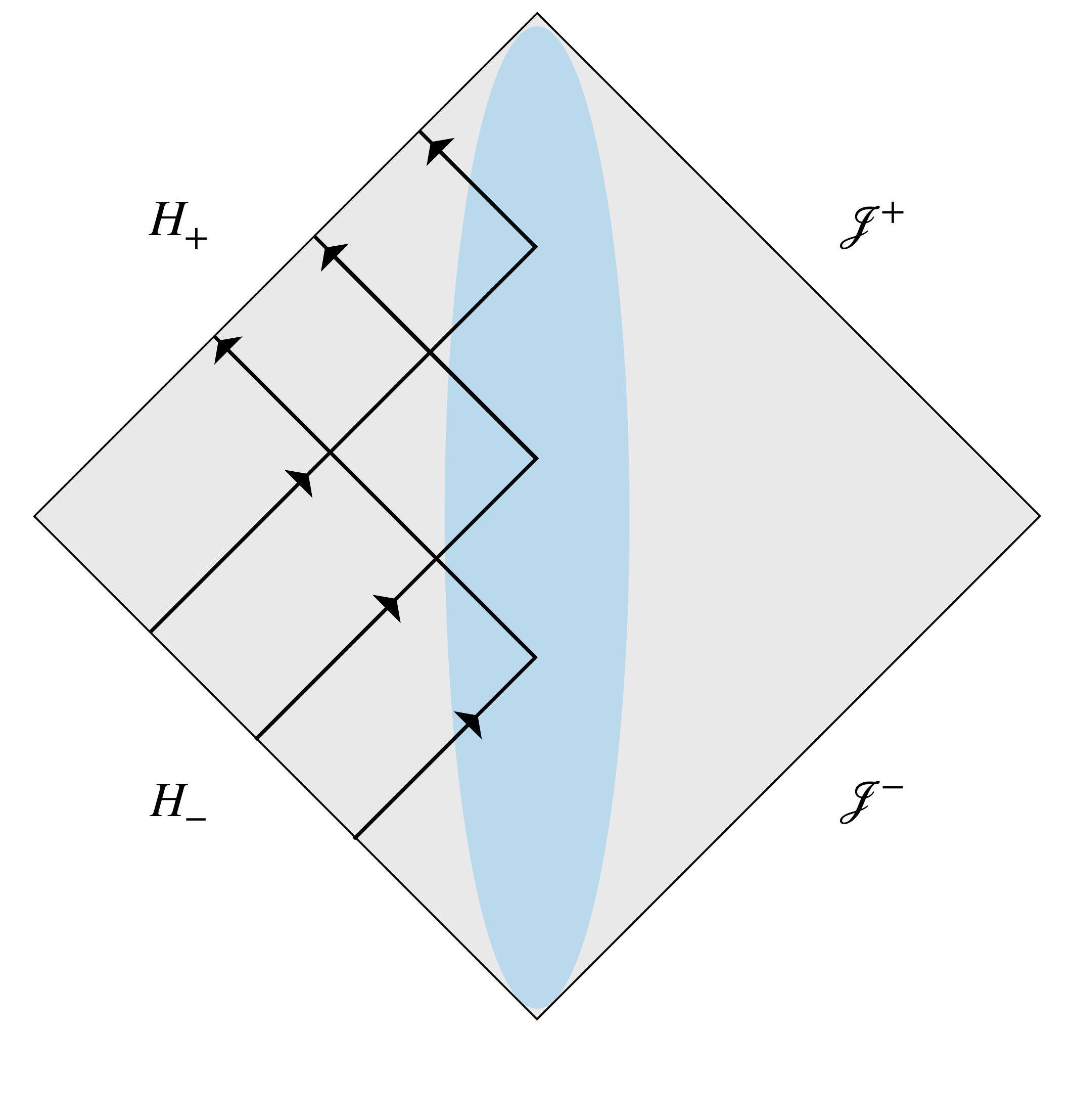}\\[-20pt]
    \caption{Principal chiral currents for $B$-state.}
    \label{Fig_3}
\end{figure}

Let us put $R_+=R_-=0$, and consider a current
\ba\label{Bstate}
\ts{J}_{B}^{\mu}= \frac{1}{ 2}(\ts{J}_+  + \ts{J}_-)=\ts{J}_{(r)}\, .
\ea
For this state there are no fluxes at the infinities ${\cal J}^{\pm}$. However, the corresponding current is singular at both (past and future) horizons. Principal chiral currents for the $B$-state are schematically shown in figure~\ref{Fig_3}.

\subsubsection{$H$-state}

\begin{figure}[!htb]%
    \centering
    \includegraphics[width=0.25\textwidth]{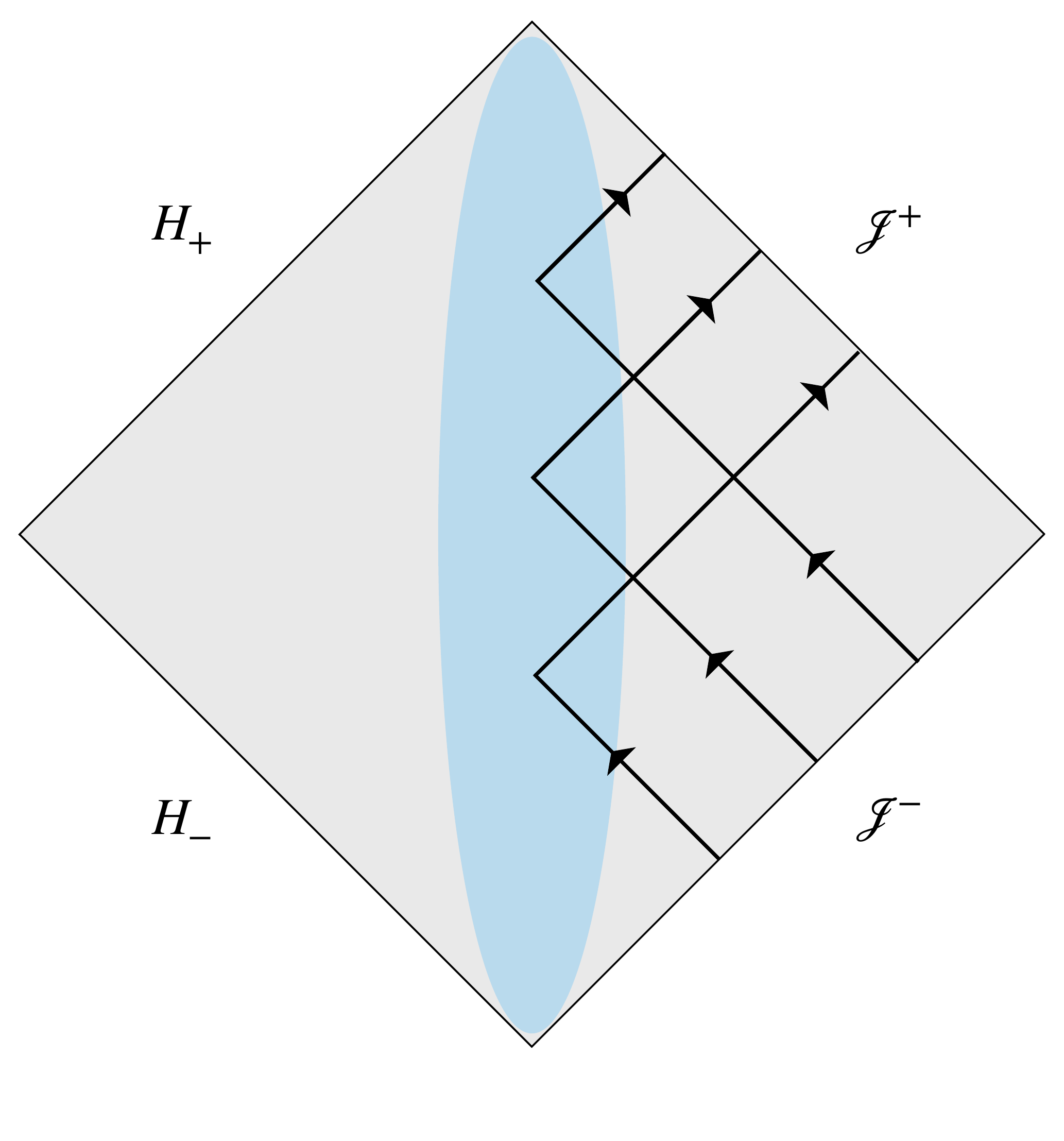}\\[-20pt]
    \caption{Principal chiral currents for $H$-state.}
    \label{Fig_4}
\end{figure}

Let us put $R_+=R_-=R_H$ and consider a current
\ba\label{Hstate}
\ts{J}_{H}= \frac{1}{ 2}(\ts{J}_+ + \ts{J}_-)=\frac{R-R_H}{ \sqrt{\Sigma\Delta_r}}e_{\bbi{1}}\, .
\ea
For this state there exists an incoming chirality flux proportional to $R_H$ at ${\cal J}^-$, which is accompanied by an outgoing chirality flux proportional to $R_H$ at ${\cal J}^+$. At both future and past horizons the chirality fluxes vanish.
Principal chiral currents for the $H$-state are schematically shown in figure~\ref{Fig_4}.

\subsubsection{$U$-state}

\begin{figure}[!htb]%
    \centering
    \includegraphics[width=0.25\textwidth]{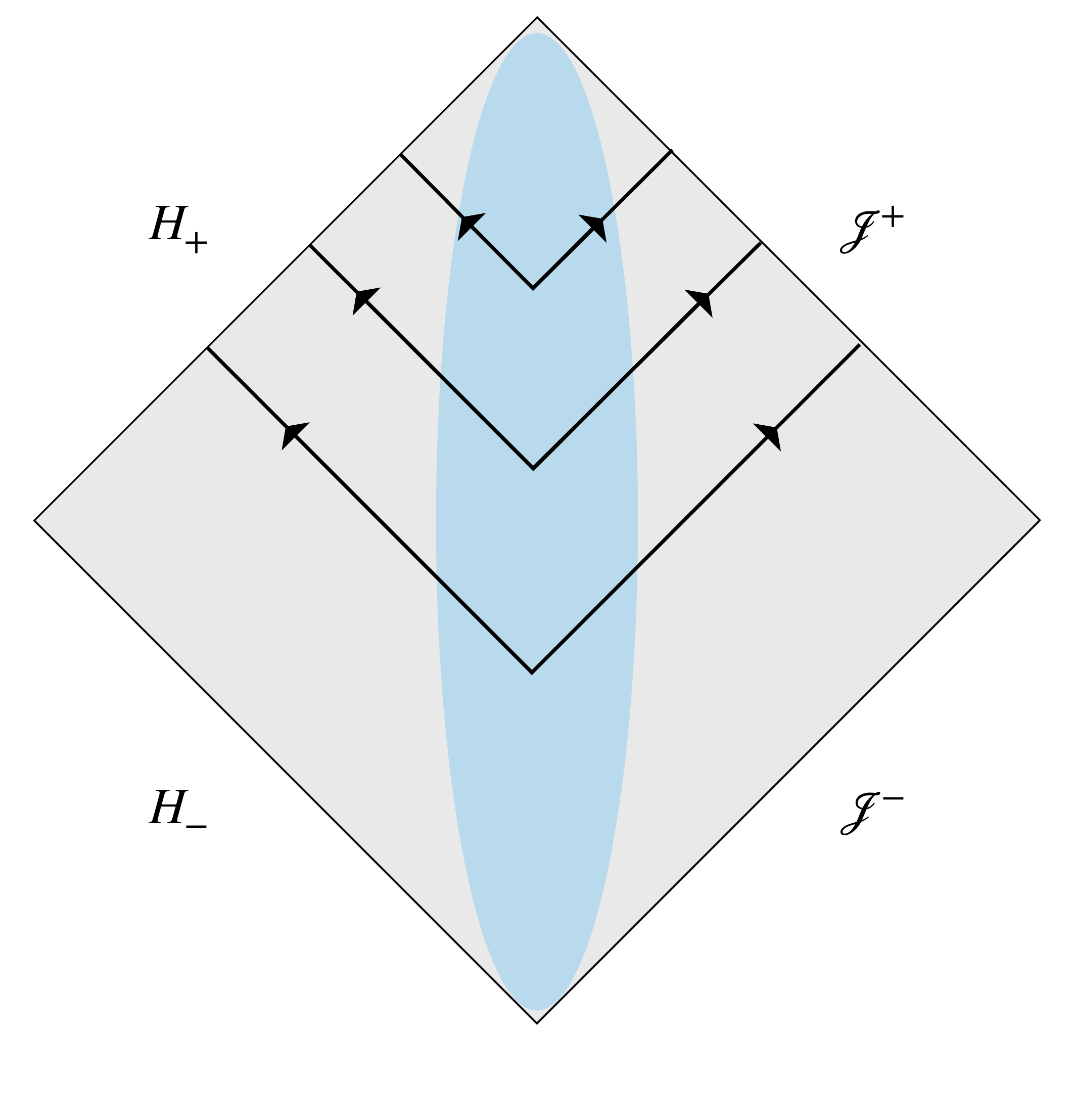}\\[-20pt]
    \caption{Principal chiral currents for $U$-state.}
    \label{Fig_5}
\end{figure}

Let us put $R_-=R_H$ and $R_+=0$ and consider a current
\ba\label{Ustate}
\ts{J}_{U}= \frac{1}{ 2}(\ts{J}_+ +\ts{J}_-)\, .
\ea
For this state there is no incoming chirality flux from the past infinity ${\cal J}^{-}$, and the current vanishes at $H_-$.   For these initial conditions, the principal chiral current contains an outgoing null flux at ${\cal J}^{+}$.  The chiral flux  at ${\cal J}^{+}$ is given by
\be\label{JU}
\ts{J}_U\sim -\frac{R_H}{ 2 r^2}\ts{l}_{-}\, .
\ee
For this state there also exists a chirality flux through the horizon $H_+$
\be
\ts{J}_U|_{H_+}=-\frac{R_H}{ 2(r_H^2+y^2)}\ts{l}_{+}\, .
\ee
Principal chiral currents for the $U$-state are schematically shown in figure~\ref{Fig_5}.

The described choices of state resemble well-known Boulware, Hartle-Hawking, and Unruh states for quantum evaporating black holes (see e.g. \cite{Candelas:1980zt,Frolov:1998wf,Casals:2012es}). This explains our notations for the states.

The property (\ref{SYMRY}) of the function $R$ implies that the chiral currents at infinity and at the horizons are antisymmetric functions of $y$. This implies that the total flux of the chirality through a 2D spherical surface at infinity vanishes. This means that the total number of particles with opposite chirality emitted by the black hole are the same. However, the integral of the flux  over a northern or southern semi-sphere does not vanish. This reflects the angular asymmetry in the emission of particles with opposite chirality, i.e. a chiral anomaly.

\subsection{Special cases}\label{sec4C}

\subsubsection{Kerr black hole}

General expressions (\ref{JJrrr}) and (\ref{RRYY})  for the currents $\ts{J}_{(r)}$ and $\ts{J}_{(y)}$ are greatly simplified for on-shell metrics. Let us consider the case of an isolated rotating black hole in an asymptotically flat spacetime.
In this case, the metric (\ref{ds2}) reduces to the Kerr metric, which has two parameters:  mass $m$ and angular momentum $ma$. One has
\ba \n{BHST}
&\Delta_r=r^2-2mr+a^2\hh \Delta_y=a^2-y^2 .
\ea
The Pontryagin invariant and functions $R$ and $Y$, which enter the expression (\ref{JJrrr}) for the chiral currents, take the form
\ba\n{PPPPRY}
&\mathcal{P}=-\frac{48m^2 ry}{(r^2+y^2)^6} (r^2-3y^2)(3r^2-y^2)\, ,\\
&R=R^m\equiv \frac{4m^2 y}{ (r^2+y^2)^4}(9r^4-14r^2y^2+y^4)\, ,\\
&Y=Y^m\equiv \frac{4m^2 r}{ (r^2+y^2)^4}(r^4-14r^2y^2+9y^4)\, .
\ea
It is interesting that the mass $m$ enters all of these expressions only in the form of prefactors proportional to $m^2$.  The expression for $\mathcal{P}$ is invariant under the change of the coordinates $r\to y$, $y\to r$. Under this transformation, $R^m\to Y^m$ and $Y^m\to R^m$.

In order to characterize properties of the Pontryagin invariant, it is instructive to consider its value at the horizon
\be\n{PHOR}
\mathcal{P}_H=-\frac{48m^2 r_H y}{(r_H^2+y^2)^6} (r_H^2-3y^2)(3r_H^2-y^2)\, .
\ee
It is easy to check that the sign of $\mathcal{P}_H$ is determined by the factor $-y(r_H^2-3y^2)$. In the northern hemisphere, where $y>0$, it anticorrelates with the sign of the Gaussian curvature $K$ of the 2D surface of the horizon. $K$ has the form \cite{Smarr:1973zz,FrolovZelnikov:2011}
\be \n{KK}
K=\frac{r_H^2+a^2}{ (r_H^2+y^2)^3} (r_H^2-3y^2)\, .
\ee
Evidently in the southern hemisphere, the sign of $\mathcal{P}_H$ correlates with that of $K$.
By comparing (\ref{PHOR}) and (\ref{KK}), one can conclude that the Pontryagin invariant on the horizon changes its sign at the equator $y=0$, and for rapidly rotating black holes it also changes sign at $|y|=r_H/\sqrt{3}$, where the Gaussian curvature $K$ vanishes. The domains with negative Gaussian curvature exist on the horizon near the poles for rapidly rotating black holes. Specifically, this occurs when their rotation parameter satisfies the inequality $a>\sqrt{3}m/2$ \cite{Smarr:1973zz}.

\begin{figure}[!htb]%
    \centering
    \includegraphics[width=0.5\textwidth]{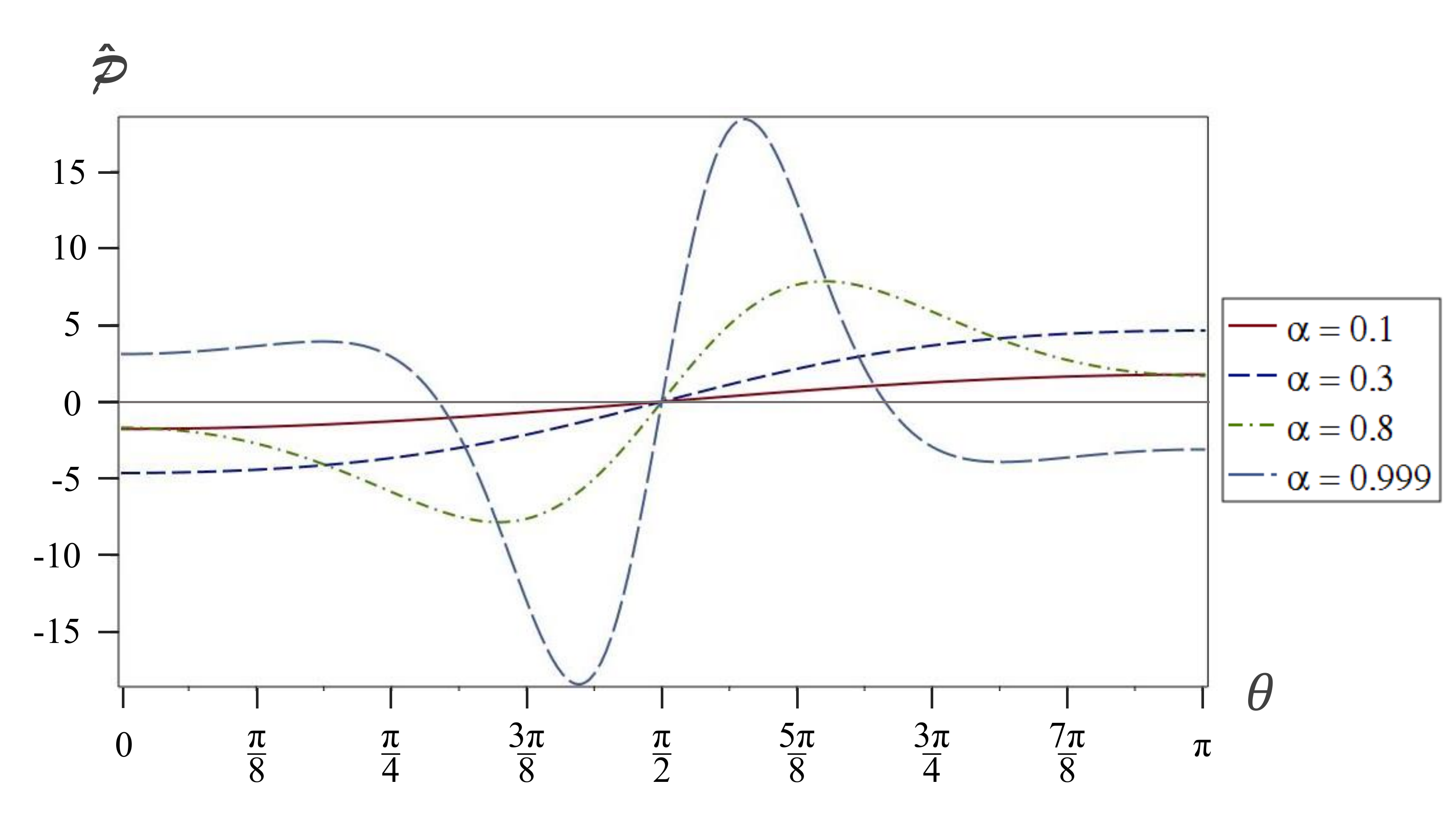}\\[0pt]
    \caption{Dimensionless Pontryagin invariant $\hat{\mathcal{P}}$ at the horizon of the Kerr black hole as a function of the angle $\theta$ for the values of the rotation parameter $\alpha=0.1, 0.3, 0.8$, and $0.999$. }
    \label{Fig_6}
\end{figure}

The Pontryagin invariant has dimensions of $[length]^{-4}$. It is convenient to define the following dimensionless version of this invariant, calculated at the horizon
\ba
&\mathcal{P}|_H=\frac{1}{r_H^4} \hat{\mathcal{P}}\, ,\\
&\hat{\mathcal{P}}=-\frac{48\alpha \rho^5\cos\theta}{ (\rho^2+\alpha^2\cos^2\theta)^6}(3\rho^2-\alpha^2\cos^2\theta)(\rho^2-3\alpha^2\cos^2\theta)\, ,\\
&\alpha=a/m\, ,\ \rho=1+\sqrt{1-\alpha^2}\, , \  y=m\alpha\cos\theta\, .
\ea
The dimensionless invariant $\hat{\mathcal{P}}$ depends on two  parameters: the dimensionless rotation parameter $0\le\alpha< 1$, and the angle $\theta$, which changes in the interval from $\theta=0$ (at the ``north pole") to $\theta=\pi$ (at the ``south pole"). Figure~\ref{Fig_6} shows the value of the dimensionless Pontryagin invariant $\hat{\mathcal{P}}$ at the horizon of the Kerr black hole as a function of the angle $\theta$.
Taking the function $R^m$ at the horizon $r=r_H=m+\sqrt{m^2-a^2}$ and the function $Y^m$ at the symmetry axes $y=\pm a$, one gets
\ba
R^m_H&=\frac{4m^2y}{(r_H^2+y^2)^4}(9r_H^4-14r_H^2y^2+y^4)\, ,\\
Y^m_A&=\frac{4m^2 r}{(r^2+a^2)^4} (r^4-14r^2a^2+9a^4)\, .
\ea
For the Kerr black hole, the chiral current in the $U$-state at infinity is
\ba\label{JU1}
&\ts{J}_U \sim \frac{1}{ m} \frac{{\cal R}_U}{ r^2}\ts{l}_- \, ,\\
&{\cal R}_U=-\frac{2\alpha \cos\theta (9\rho^4-14\rho^2\alpha^2\cos^2\theta+\alpha^4\cos^4\theta)}{ (\rho^2+\alpha^2\cos^2\theta)^4}\, .
\ea
Note that this current was normalized to satisfy Eq.(\ref{MAIN}), while the quantum average of the current for the massless Dirac neutrino field (with the spin $s=1/2$) obeys Eq.(\ref{FERM}), which differs from $\ts{J}_U$ by a factor of $(96\pi^2)^{-1}$. In the case of electromagnetic field (spin $s=1$) this factor would be $(48\pi^2)^{-1}$. Thus, the chirality flux density for the quantum field of the spin $s$ is obtained by the substitution ${\cal R}_U \to {\cal R}$, where
\ba\label{RRU}
{\cal R}=\frac{n(s)}{96\pi^2}{\cal R}_U
\ea
and the coefficient $n(s)$ depends on the spin of the quantum field, so that $n(1/2)=1$ and $n(1)=2$.

\begin{figure}[!htb]%
    \centering
    \includegraphics[width=0.48\textwidth]{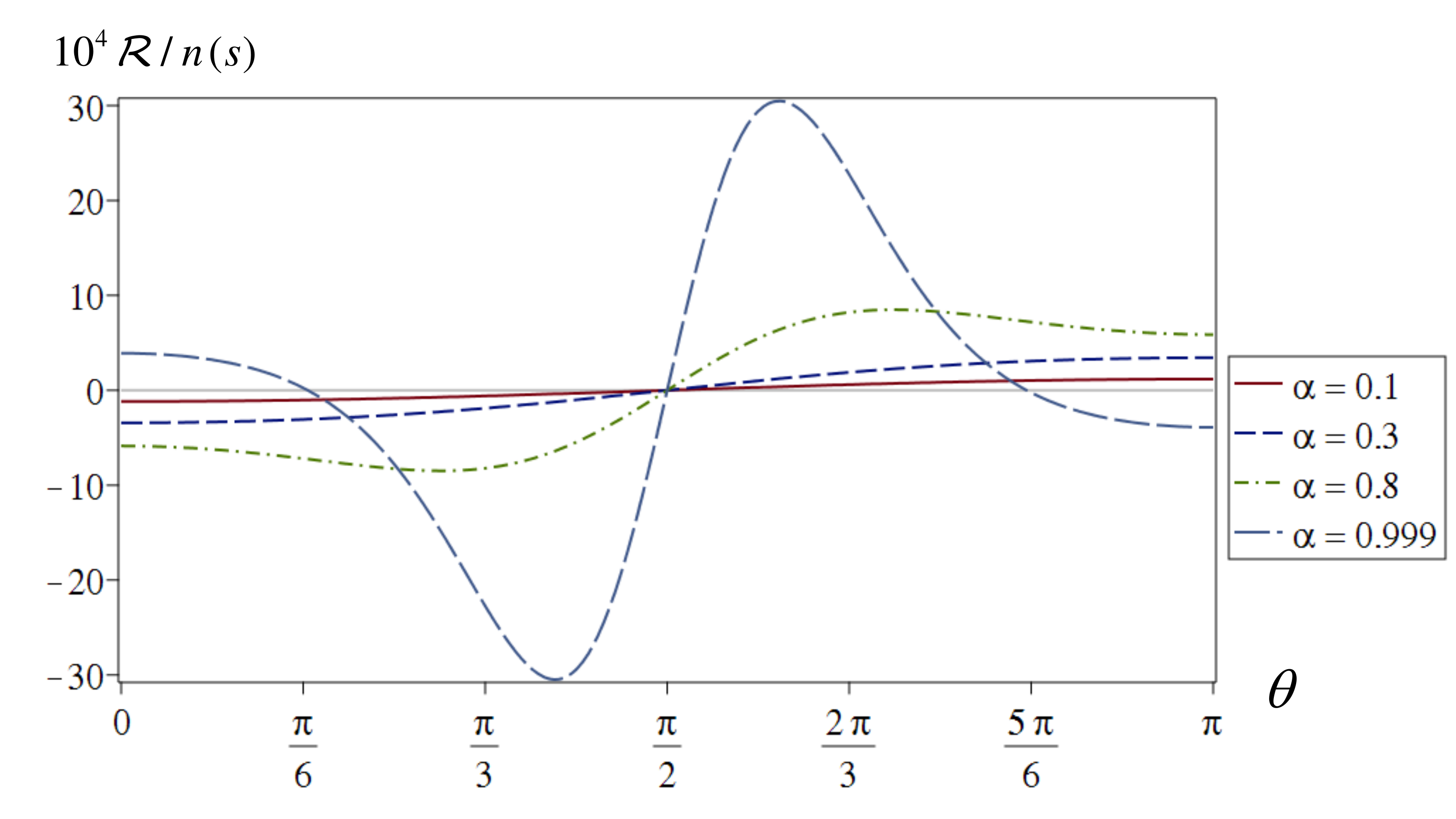}\\[0pt]
    \caption{Dimensionless chiral current flux at infinity ${\cal R}$ for the Kerr black hole as a function of the angle $\theta$ for the values of the rotation parameter $\alpha=0.1, 0.3, 0.8$, and $0.999$. }
    \label{Fig_7}
\end{figure}

Figure~\ref{Fig_7} shows plots of the dimensionless quantity ${\cal R}$ as a function of the angle $\theta$ for different values of the dimensionless rotation parameter.

Since the gravitational Pontryagin density $\mathcal{P}$  is an antisymmetric function of the angle variable $y$, the corresponding total flux of the chiral current calculated for a 2D spherical surface surrounding the black hole vanishes. One can define a quantity that characterizes the current flux which takes into account this asymmetry. This can be done as follows. Let us write the current $\ts{J}_U$ in the $U$-state in the (3+1)-form $J_U^{\mu}=(-\rho,\vec{j})$.
The component of the current which lies in the direction of the black hole's axis of rotation is $j^z=j^r \cos\theta$. Using the relation $y=a\cos\theta$, and after rescaling (\ref{RRU}) and taking the integral of $j^z$ over the surface of a sphere ($r$=const) as $r\to\infty$, one gets
\be\n{LLLL}
\dot{L}_m=\frac{n(s)}{96\pi^2}\lim_{r\to\infty} \left[ \frac{r^2}{ a^2}\int_0^{2\pi} d\phi \int_{-a}^a dy y\, j^r\right]\, .
\ee
The quantity $\dot{L}_m$ describes the loss of angular momentum in the rotating black hole due to the spin of radiated chiral particles.

Calculating the integral in (\ref{LLLL}), we obtain
\ba \n{LLmm}
\dot{L}_m&=-n(s)\frac{a m^2(3 r_H^2-a^2)}{12\pi (r_H^2+a^2)^3}\\
&=-n(s)\frac{1}{12}A_H T_H^2\Omega_H\frac{m^2(3 r_H^2-a^2)}{(r_H-m)^2(r_H^2+a^2)} ,
\ea
where we took into account that the angular velocity of the Kerr black hole $\Omega_H$, the surface area of the horizon $A_H$, and Hawking temperature are
\ba\label{OmegaH}
&\Omega_H=\frac{a}{r_H^2+a^2} \hh
A_H=4\pi(r_H^2+a^2),\\
&T_H=\frac{r_H-m}{2\pi (r_H^2+a^2)}.
\ea
In the limit of a slowly rotating black hole $\alpha=a/m \ll 1$, we get
\ba \n{LLmm0}
\dot{L}_m& \simeq -n(s)\frac{\Omega_H}{16\pi} .
\ea

\begin{figure}[!htb]%
    \centering
    \includegraphics[width=0.4\textwidth]{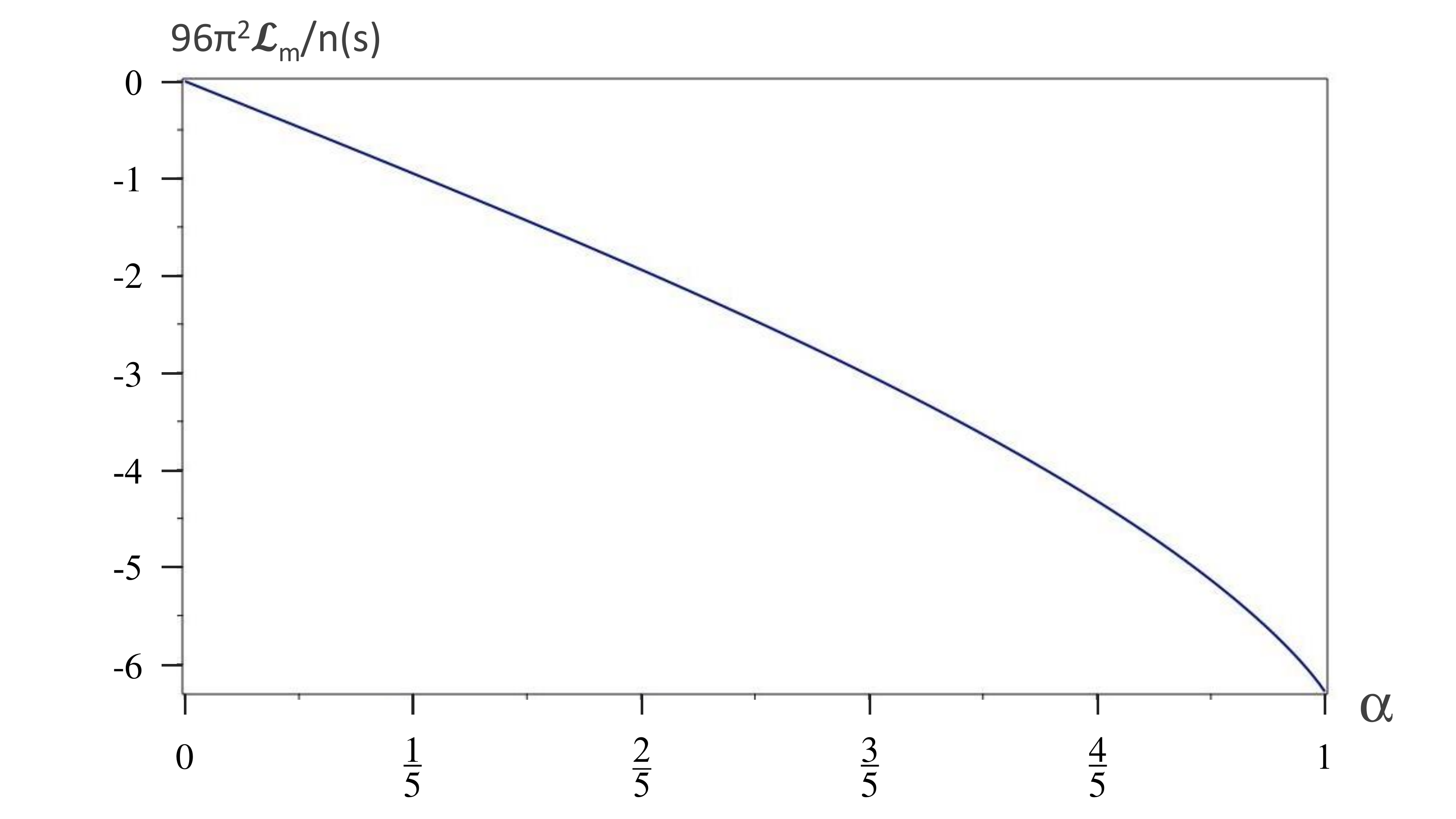}\\[0pt]
    \caption{${\cal L}_m$ as a function of the dimensionless rotation parameter $\alpha$. }
    \label{Fig_8}
\end{figure}

It is convenient to write the expression (\ref{LLmm}) in the dimensionless form
\ba
&\dot{L}_m=\frac{1}{ m}{\cal L}_m\, ,\\
&{\cal L}_m=-n(s)\frac{\alpha(3\rho^2-\alpha^2)}{12\pi (\rho^2+\alpha^2)^3}\, .
\ea
Here ${\cal L}_m$ is a dimensionless function of the dimensionless rotation parameter $\alpha$.
A plot of ${\cal L}_m$ as a function of the dimensionless rotation parameter $\alpha$ is given in Fig.\,\ref{Fig_8}.

\subsubsection{Kerr-Newman black hole}

If a rotating black hole has either electric charge $Q$, or magnetic monopole charge $P$, or both, then the off-shell metric (\ref{ds2}) takes the Kerr-Newman form for which $\Delta_y$ remains the same as in the uncharged case, while $\Delta_r$ becomes
\ba \n{BHST}
&\Delta_r=r^2-2mr+a^2+\widetilde{Q}^2\hh \widetilde{Q}^2=Q^2+P^2, .
\ea
 The Pontryagin invariant for this metric is
\ba\n{MMPP}
\mathcal{P}=&\frac{48y}{(r^2+y^2)^6}[mr(r^2-3y^2)-\widetilde{Q}^2(r^2-y^2)]\\
&\times[m(y^2-3r^2)+2r\widetilde{Q}^2]\, .
\ea
Substituting the expressions for $\Delta_r$ and $\Delta_y$ into (\ref{JJrrr}) gives the corresponding expressions for the functions $R$ and $Y$. It is interesting that both of these functions can be written as the sum of two terms
\be \n{RRQQYY}
R=R^m+R^Q\hh Y=Y^m+Y^Q\, ,
\ee
where $R^m$ and $Y^m$ are the functions $R$ and $Y$ calculated above for the Kerr metric, while
$R^Q$ and $Y^Q$ depend on charge, and are of the form
\ba
R^Q=&\frac{8y\widetilde{Q}^2}{(r^2+y^2)^4}[\widetilde{Q}^2 (2r^2-y^2)-6mr(r^2-y^2)]\, ,\\
Y^Q=&\frac{4\widetilde{Q}^2}{(r^2+y^2)^4}[2\widetilde{Q}^2 r(r^2-2y^2)\\
&-3m(r^4-6r^2y^2+y^4)]  .
\ea

For the Kerr-Newman metric
\be
r_H=m+\sqrt{m^2-a^2-\widetilde{Q}^2}\, .
\ee
Using the above expressions (\ref{RRQQYY}), one can find  $R_H$ and $Y_A$.
Calculating the integral (\ref{LLLL}), one obtains the following expression
\ba
&\dot{L}=\dot{L}_m+\dot{L}_Q\, ,\\
&\dot{L}_Q=n(s)\frac{\widetilde{Q}^2}{96\pi r_H^3 a^2}\Big\{\frac{32m  a^3 r_H^4}{(a^2+r_H^2)^3}\\
&-\widetilde{Q}^2\Big[\arctan\Big(\frac{a}{r_H}\Big)+\frac{a r_H( a^4+8 a^2 r_H^2- r_H^4)}{(a^2+r_H^2)^3}\Big]\Big\} .
\ea
Here $\dot{L}_m$ is given by (\ref{LLmm}).

\section{Chern-Simons chiral current}
\label{s5}
\subsection{General expression of the Chern-Simons current for the off-shell metric}

In the previous section, we discussed special solutions of the chiral anomaly equations. We obtained an explicit form of the solutions for the chiral currents, which have special symmetry properties. Let us now discuss another quite general approach for solving the chiral anomaly equations. Let $\ts{e}_A$ be a normalized basis, and let the index $A$ enumerate the basic vectors. Let us use it to define the Chern-Simons current $\ts{I}$
 \be
I^\alpha=-\frac{1}{2}e^{\alpha\beta\mu\nu}\left(R_{\beta\mu AB}\,\omega_{\nu}{}^{AB}
+\frac{2}{ 3}\,\omega_{\beta}{}^{A}{}_{B}\,\omega_{\mu}{}^{B}{}_{C}\,\omega_{\nu}{}^{C}{}_{A}\right)\, .
\ee
Here
\ba
&\omega_{\mu}^{AB}=e_{\nu}^A\nabla_{\mu}e^{\nu B}\, ,\\
&R_{\mu\nu AB}=R_{\mu\nu\alpha\beta}e^\alpha_Ae^\beta_B\, .
\ea
One can show that this current obeys the equation
\be
I^{\mu}_{\ ;\mu}=\mathcal{P}\, ,
\ee
where $\mathcal{P}$ is the Pontryagin invariant. (For details and general discussion see e.g. \cite{Eguchi:1980jx}).
In the general case, the Chern-Simons current depends on the choice of basis.

In the case we discussed in this paper, explicit and hidden symmetries of the off-shell metric (\ref{ds2}) single out a special Darboux tetrad, for which properties of the metric are greatly simplified. Using the Darboux basis
$e^\alpha_A=(e_1^\alpha,e_{\bar{1}}^\alpha,e_2^\alpha,e_{\bar{2}}^\alpha)$ defined by  (\ref{DDEE}), and after quite long but straightforward calculations, one obtains \footnote{The current $\ts{I}$ for the off-shell metric in Darboux coordinates also has
the following explicit form in terms of Christoffel symbols
$
I^\alpha=e^{\alpha\beta\mu\nu}\Big(\Gamma_{\beta\lambda}^{\sigma} \partial_{\mu} \Gamma_{\nu\sigma}^{\lambda}
+\frac{2}{3}\Gamma_{\beta\lambda}^{\sigma}\Gamma_{\mu\epsilon}^{\lambda}\Gamma_{\nu\sigma}^{\epsilon}
\Big)
$
\cite{Miskovic:2009bm}  .
}
\ba\label{Jr1}
I^{\mu}=&(0,I^r,I^y,0)\, ,\\
I^r=&\frac{1}{\Sigma^5}\Big\{
y\Sigma^2 \big[\Delta_{r}\ddot{\Delta}_y+(\Delta_{r}')^2\big]\\
&-r\Sigma\Delta_{r}'\big[\Sigma\dot{\Delta}_y+4y(2\Delta_{r}-\Delta_{y})
\big]  \\
&+\Sigma(3\Sigma-8y^2)\Delta_{r}\dot{\Delta}_y\\
&+8y(2\Sigma-3y^2)\Delta_{r}(\Delta_{r}-\Delta_{y})
\Big\}\, ,\\
I^y=&\frac{1}{\Sigma^5}\Big\{-
r\Sigma^2 \big[\Delta_{y}\Delta_{r}''+(\dot{\Delta}_y)^2\big]\\
&+y\Sigma\dot{\Delta}_y\big[\Sigma\Delta_{r}'-4r(\Delta_{r}-2\Delta_{y})
\big]  \\
&+\Sigma(5\Sigma-8y^2)\Delta_{y}\Delta_{r}'\\
&-8r(\Sigma-3y^2)\Delta_{y}(\Delta_{r}-\Delta_{y})
\Big\}
\ea

Using these general expressions, one can calculate the functions $R_H$ on the horizon and $Y_A$ on the axis of symmetry. One has
\ba\n{AAA}
R_H&=\frac{\Delta_r'}{\Sigma^3} \left(
\Sigma(y\Delta_r'-r\dot{\Delta}_y)+4ry\Delta_y
\right)\Big|_{r=r_H} ,\\
Y_A&=\frac{\dot{\Delta}_y}{\Sigma^3}\left(
\Sigma(y\Delta_r'-r\dot{\Delta}_y)-4ry \Delta_r
\right)\Big|_{y=a}.
\ea

\subsection{Chern-Simons current in the Kerr spacetime}\label{sec5B}

For the on-shell Kerr metric, the functions $R_H$ and $Y_A$ take the form
\ba\n{AAA1}
R_H=&\frac{4my(r_H-m)(3r_H^2-y^2)}{(r_H^2+y^2)^3},\\
Y_A=&-\frac{4ma^2(3r^2-a^2)}{(r^2+a^2)^3} .
\ea

The Chern-Simons chiral anomaly current $\ts{I}$ is an $S$-vector, and its temporal component vanishes. One can ``upgrade" this current to include fluxes at infinity and the horizon by adding a corresponding homogeneous solution, as it was done in the previous subsection for the principal current. Namely, we define two currents $\ts{I}_{\pm}$
\be
I^{\mu}_{\pm}=\mp (I^r-\frac{R_{\pm}}{ \Sigma}) l_{\pm}^{\mu}+(I^y-\frac{Y_A}{\Sigma})\delta_y^{\mu}\, .
\ee
The term $R_{\pm}$ depends on the choice of the state. For the $U$-state, one has
\be
I^{\mu}_{U}=  \frac{1}{ 2}\left[ (I^r-\frac{R_{H}}{ \Sigma}) l^{\mu}_{-}-I^r l^{\mu}_{+}
\right]
+({I^y}-\frac{Y_A}{ \Sigma})\delta_y^{\mu}\, .
\ee
For the $U$-state, the upgraded Chern-Simons current describes the following chirality flux at spatial infinity
\ba\label{CS}
&{\ts{I}_U \sim }\frac{1}{ m}\frac{ {\cal R}_U}{ r^2}\ts{l}_- \, ,\\
&{\cal R}_U=-\frac{2\alpha\cos\theta}{ (\rho^2+\alpha^2\cos^2\theta)^3}(\rho-1)(3\rho^2-\alpha^2\cos^2\theta)\, .
\ea
Calculating the integral in (\ref{LLLL}) for the Chern-Simons current, we obtain the flux
\ba \n{LLCS}
\dot{L}_m&=-n(s)\frac{a m(r_H-m)}{12\pi (r_H^2+a^2)^2} .
\ea
It is instructive to rewrite this result in terms of the black hole angular velocity, the surface area of the horizon, and the Hawking temperature (\ref{OmegaH})
\ba \n{LLCS1}
\dot{L}_m&=-n(s)\frac{1}{12}A_H T_H^2\Omega_H\Big(\frac{m}{r_H-m}\Big)  .
\ea

In the limit of a slowly rotating black hole, we get
\ba \n{LLCS0}
\dot{L}_m& \simeq -n(s)\frac{\Omega_H}{48\pi} .
\ea

\begin{figure}[!htb]%
    \centering
    \includegraphics[width=0.48\textwidth]{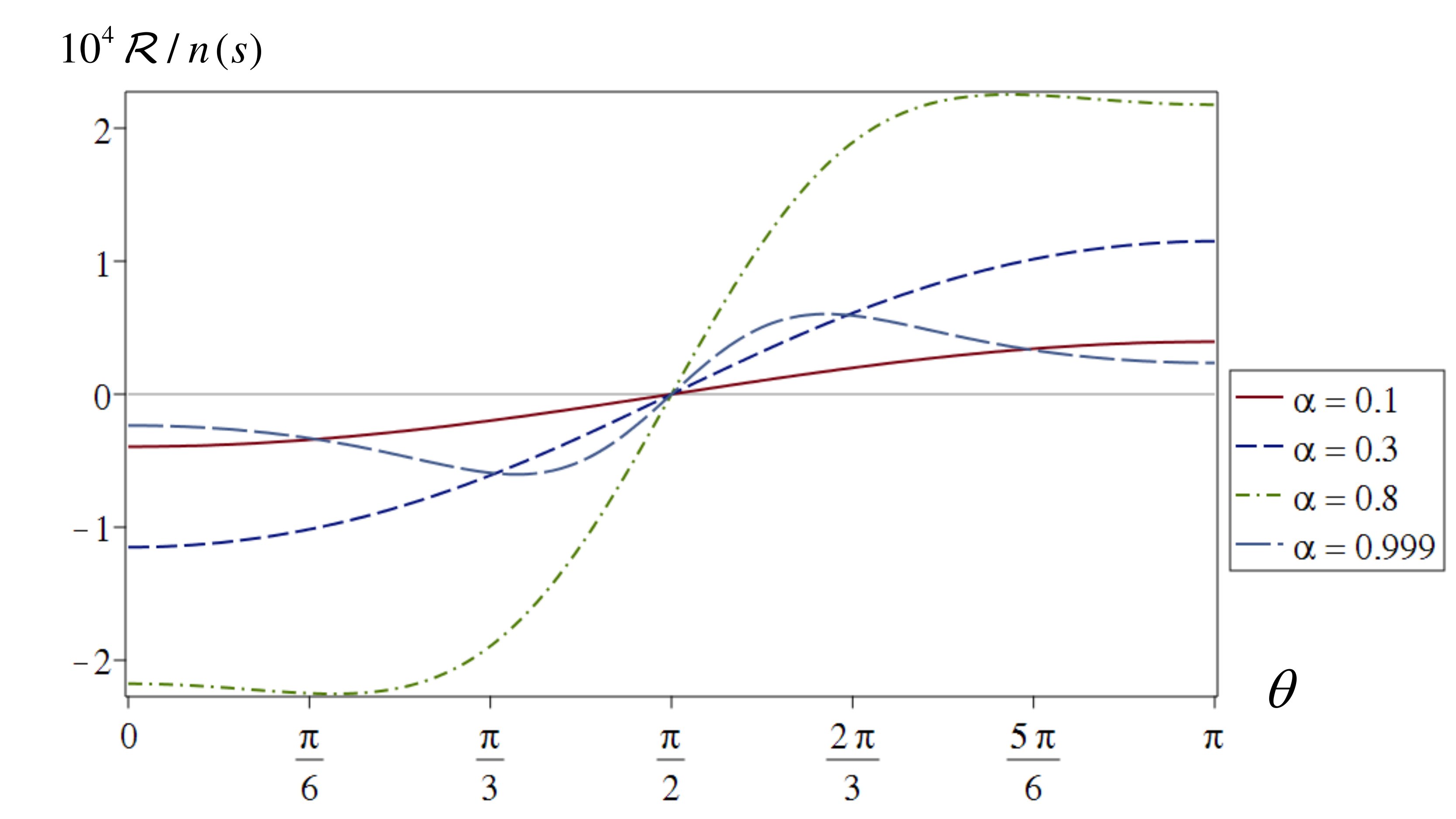}\\[0pt]
    \caption{Dimensionless Chern-Simons flux  at infinity ${\cal R}$ for the Kerr black hole as a function of the angle $\theta$ for the values of the rotation parameter $\alpha=0.1, 0.3, 0.8$, and $0.999$. }
    \label{Fig_9}
\end{figure}

Figure~\ref{Fig_9} shows plots for the dimensionless quantity ${\cal R}$ as a function of the angle $\theta$ for different values of the dimensionless rotation parameter $0\le\alpha < 1$.

Let us note that in paper \cite{Flachi:2017vlp}, the authors suggested using the Chern-Simons current for evaluating the chiral current of a slowly rotating black hole with angular velocity $\Omega$. They arrived at the conclusion that the leading-order in $\Omega$ contribution to the chirality flux vanishes quickly at large distance, and it does not contribute to the flux at infinity. This result can be explained as follows. The expression for the chiral current, which was used by these authors, does not satisfy the regularity condition at the horizon, and in this sense it is similar to the choice of the Boulware vacuum state.

\subsection{Relation between Chern-Simons currents and the principal currents}

Let us note that we have already found two special expressions for the chiral currents $\ts{J}_{(r)}$ and $\ts{J}_{(y)}$. Each of these currents is a solution of the inhomogeneous chiral anomaly equation (\ref{MAIN}) respecting both explicit and hidden symmetries of the spacetime for metric (\ref{ds2}). This means that the Chern-Simons current $\ts{I}$ can be presented in the following two forms
\be\label{IZ}
I^{\mu}=J_{(r)}^{\mu}+Z_{(r)}^{\mu}\hh I^{\mu}=J_{(y)}^{\mu}+Z_{(y)}^{\mu}, .
\ee
Both of the vectors $\ts{Z}_{(r)}$ and $\ts{Z}_{(y)}$ are solutions to the homogeneous equation
\be
Z_{(r)\ ;\mu}^{\mu}=Z_{(y)\ ;\mu}^{\mu}=0\, .
\ee
They are $S$-vectors, with the following components
\ba
Z_{(r)}^{\mu}&=(0,Z_{(r)}^r,Z_{(r)}^y,0)\, ,\\
Z_{(y)}^{\mu}&=(0,Z_{(y)}^r,Z_{(y)}^y,0)\, ,\\
Z_{(r)}^r&=\frac{1}{\Sigma^5}\Big\{+r\Sigma^2\Delta_{r}'\dot{\Delta}_y -4 ry\Sigma\Delta_{r}'\Delta_{y} \\
&-\Sigma^2\ddot{\Delta}_y \big[\Sigma\dot{\Delta}_y+y(\Delta_{r}-2\Delta_{y})\big]\\
&+\Sigma(8y^2-3\Sigma)(\Delta_{r}-2\Delta_{y})\dot{\Delta}_y +3y\Sigma^2 (\dot{\Delta}_y)^2 \\
&   -8y(3y^2-2\Sigma)(\Delta_{r}-\Delta_{y})\Delta_{y}
\Big\}  ,\\
Z_{(r)}^y&=\frac{1}{\Sigma^5}\Big\{\big[y\Sigma^2\dot{\Delta}_y+\Sigma(8y^2-5\Sigma)\Delta_{y}  \big]\Delta_{r}'\\
&-r\Sigma^2(\dot{\Delta}_y)^2-4ry\Sigma(\Delta_{r}-2\Delta_{y})\dot{\Delta}_y\\
&-r\Sigma^2\Delta_{y}\Delta_{r}''+8r (3y^2-\Sigma)\Delta_{y}(\Delta_{r}-\Delta_{y})
\Big\}  ,
\ea
\ba
Z_{(y)}^r&=\frac{1}{\Sigma^5}\Big\{-r\Sigma\big[\Sigma\dot{\Delta}_{y} +4y(2\Delta_r-\Delta_y)\big]\Delta_{r}'\\
&-\Sigma(8y^2-3\Sigma)\Delta_r\dot{\Delta}_{y}+y\Sigma^2\Delta_r\ddot{\Delta}_{y}\\
&-8y(3y^2-2\Sigma)\Delta_r(\Delta_r-\Delta_y)+y\Sigma^2(\Delta_{r}')^2
\Big\}  ,\\
Z_{(y)}^y&=\frac{1}{\Sigma^5}\Big\{\Sigma\big[-\Sigma y\dot{\Delta}_{y} +\Sigma^2\Delta_{r}'' \\ &-(8y^2-5\Sigma)(2\Delta_r-\Delta_y)\big]\Delta_{r}'\\
&+4ry\Sigma\Delta_r\dot{\Delta}_{y}-r\Sigma^2 (2\Delta_r-\Delta_y)\Delta_{r}''\\
&+8r(3y^2-\Sigma)\Delta_r(\Delta_r-\Delta_y)-3r\Sigma^2(\Delta_{r}')^2
\Big\}  .
\ea

Since the currents $\ts{Z}_{(r)}$ and $\ts{Z}_{(y)}$ are conserved, there exist scalar potentials $\Psi_{(r)}$ and $\Psi_{(y)}$  which generate these currents
\ba
Z_{(i)}^{\mu}=-\frac{1}{\Delta_r\Delta_y}e^{\mu\nu\rho\sigma}\Psi_{(i)}{}_{;\nu}\xx{0}{}_{\rho}\xx{1}{}_{\sigma}\, .
\ea
Here $i=r,y$. These potential are
\ba
\Psi_{(r)}=&\frac{1}{2\Sigma^3}\Big[
\Sigma^2(\dot{\Delta}_{y})^2+2y\Sigma(\Delta_r-2\Delta_y)\dot{\Delta}_{y} \\
&+2\Delta_y\big[
-r\Sigma \Delta_{r}'+2(\Sigma-2y^2)(\Delta_r-\Delta_y)
\big]
\Big] ,
\ea
\ba
\Psi_{(y)}=&\frac{1}{2\Sigma^3}\Big[
\Sigma^2(\Delta_{r}')^2-2r\Sigma(2\Delta_r-\Delta_y)\Delta_{r}' \\
&+2\Delta_r\big[
-y\Sigma \dot{\Delta}_{y}+2(\Sigma-2y^2)(\Delta_r-\Delta_y)
\big]
\Big] ,
\ea

In the case of the Kerr-Newman black hole, these potentials become
\ba
\Psi_{(r)}=&\frac{2}{(r^2+y^2)^3}\big\{-mr[a^2(r^2-3y^2)-y^2(3r^2-y^2)]\\
&+\widetilde{Q}^2(a^2r^2-a^2y^2-2r^2y^2)
\big\} ,
\ea
\ba
\Psi_{(y)}=&\frac{2}{(r^2+y^2)^3}\big\{-m[a^2r(r^2-3y^2)-ry^2(3r^2-y^2)\\
&-m(r^4-6r^2y^2+y^4)]\\
&+\widetilde{Q}^2[\widetilde{Q}^2(r^2-y^2)+(a^2r^2-a^2y^2-2r^2y^2)\\
&-2mr(r^2-3y^2)]
\big\} .
\ea

\section{Chiral anomaly induced by the electromagnetic field}
\label{s6}
\subsection{Electromagnetic contribution to the chiral anomaly current}

In the presence of an external electromagnetic field, the axial-current anomaly for a massless Dirac field $\psi$ with electric charge $e$ (\ref{FERM}) also contains a contribution from the invariant $F_{\mu\nu}{}^*\!F^{\mu\nu}$. When a rotating black hole has an electric and/or magnetic charge, this invariant does not vanish. Since the equation for the chiral current (\ref{FERM}) is linear, one can calculate the contribution from the external electromagnetic field independently of the curvature contribution. For this purpose, we consider first a homogeneous solution of the Maxwell equation on the background of the off-shell metric (\ref{ds2}) which respects the metric's symmetry. We specify the solution so that it describes the electromagnetic field generated by an electric charge $Q$ and magnetic monopole charge $P$. We perform these calculation first without specifying the arbitrary functions $\Delta_r(r)$ and $\Delta_y(y)$. For a special choice
(\ref{BHST}) of these functions with $b=\sqrt{m^2-a^2-Q^2-P^2}$, one reproduces the Kerr-Newman solution of the Einstein-Maxwell equations describing a rotating charged black hole.

Let us consider the following one-form for the 4D electromagnetic field potential
\be \n{POTEN}
\ts{A}=A_{\mu}dx^{\mu}=-\frac{1}{ \Sigma}\left[
Q r(d\tau+y^2 d\psi)+P y(d\tau-r^2 d\psi)
\right]\, .
\ee
It is possible to check that this potential satisfies the Lorentz gauge condition $A^{\mu}_{\ ;\mu}=0$.
The non-vanishing components of the field $\ts{F}=d\ts{A}$ are
\ba
&F_{\tau r}=- U, \hskip 1.2cm  F_{\tau y}=- V  ,\\
&F_{\psi r}=- y^2 U, \hskip 0.8cm F_{\psi y}=  r^2 V  ,
\ea
where
\ba\label{UV}
U=\frac{2Pry+Q(r^2-y^2)}{\Sigma^2} ,\\
V=\frac{2Qry-P(r^2-y^2)}{\Sigma^2} \, .
\ea
It is easy to check that this field  satisfies the homogeneous Maxwell equations
\be
F^{\mu\nu}\!{}_{;\nu}=0\, .
\ee

Direct calculations show that the non-vanishing components of the dual field  ${}^*\!\ts{F}$ are
\ba
&{}^*\!F_{\tau r}=- V   \hh \hskip 0.4cm  {}^*\!F_{\tau y}=  U,\\
& {}^*\!F_{\psi r}=- y^2 V   \hh    {}^*\!F_{\psi y}=- r^2 U   .
\ea
This field satisfies the equations
\be
{}^*\!{F}^{\mu\nu}{}_{\!;\nu}=0\, .
\ee
One also has $\ts{d}{}^*\!\ts{F}=0$, and so there exists a one-form $\ts{B}$ such that
\be
{}^*\!F_{\mu\nu}=\pa_{\mu}B_{\nu}-\pa_{\nu}B_{\mu}\, .
\ee
This potential is
\be
B_{\mu}dx^{\mu}=-\frac{1}{ \Sigma}\left[Qy(d\tau-r^2 d\psi)-Pr(d\tau+y^2 d\psi)\right]\, .
\ee
Let us note that the potential $\ts{B}$ can be obtained from $\ts{A}$ by the following transformation
\be
Q\to -P\hh P\to -Q\hh y\to -y\, .
\ee

\subsection{Principal chiral current}

Let us consider the invariant
\ba\n{TRKK}
\mathcal{P}_e&=F_{\mu\nu}{}^*\!F^{\mu\nu} .
\ea
For the potential (\ref{POTEN}) it reads
\ba
\mathcal{P}_e&=-\frac{4}{\Sigma^4} [ 2Qry-P(r^2-y^2)][Q(r^2-y^2)+2Pry]\\
  &=-4 UV
\ea

We now discuss solutions to the chiral anomaly equation (\ref{MAIN}) with right-hand side $\mathcal{P}=\mathcal{P}_e$. Like before, we first consider $S$-currents $\ts{J}_{(r)}=(0,J_{(r)}^r,0,0)$ and $\ts{J}_{(y)}=(0,0,J_{(y)}^y,0)$ which respect the hidden symmetry and are eigenvectors of the Killing tensor $\ts{K}$.
We call them the $R$-current and $Y$-current, respectively. These currents can be found by integrating the following equations
\be
\pa_r(\Sigma {J}_{(r)}^r)=\Sigma \mathcal{P}_e   \hh    \pa_y(\Sigma {J}_{(y)}^y)=\Sigma \mathcal{P}_e\, .
\ee
We write these solutions in a form similar to (\ref{JJrrr})
\ba\n{JJJrr}
&{J}^{\mu}_{(r)}=\frac{R-R_0(y)}{ \Sigma} \delta^{\mu}_r\hh
R=\int dr \,\Sigma \mathcal{P}_e\, ,\\
&{J}^{\mu}_{(y)}=\frac{Y-Y_0(r)}{ \Sigma} \delta^{\mu}_y\hh
Y=\int dy \,\Sigma \mathcal{P}_e\, .
\ea
Here
\ba\label{RY}
& R=rW     \hh    Y=-yW ,\\
&W=\frac{4}{ \Sigma^2}(Qy-Pr)(Qr+Py)\, ,
\ea
and $R_0(y)$ and $Y_0(r)$ are the corresponding ``integration constants".
Let us emphasize that both $R$ and $Y$ do not depend on a special form of the metric function $\Delta_r$, and hence they are the same for the off-shell and on-shell metrics.

Let us note that the function ${W}$ contains both symmetric and antisymmetric parts with respect to the
reflection $y\to -y$. The symmetric part, which is proportional to $QP$, gives an antisymmetric
contribution to ${J}^{\mu}_{(y)}$. Hence if $QP\ne 0$, it is impossible to choose a function  $Y_A(r)$ in (\ref{JJJrr}) which
makes the current $\bar{J}^{\mu}_{(y)}$ regular at both axes $y=\pm a$ simultaneously . In order to exclude
non-regular currents for ${J}^{\mu}_{(y)}$, one should impose the condition $QP=0$.

In what follows, we focus on the $R$-current ${\ts{J}}_{(r)}$. This current is well defined for arbitrary values of $Q$ and $P$. However, in order to simplify expressions, in what follows we assume that the magnetic
monopole charge $P$ vanishes. Thus
\ba
W=\frac{4Q^2 ry}{\Sigma^2} .
\ea

Following the procedure described in subsection (\ref{ssPCC}), one can ``upgrade" the $R$-current by introducing
incoming and outgoing principal null chiral currents
\be
{\ts{J}}_{\pm}=\mp \frac{R-R_{\pm}(y)}{ \Sigma} \ts{l}_{\pm}\, .
\ee
Using these currents, one can reconstruct the chiral fluxes for the $B$, $H$, and $U$ states (\ref{Bstate})-(\ref{Ustate}),
and then using expression (\ref{ASSJ}) one can calculate the fluxes at infinity and at the horizon.

Note that the functions $R$ and $Y$ (\ref{RY}) in the principal currents (\ref{JJJrr}) do not depend on
the functions $\Delta_r$ and $\Delta_y$  of the metric. For this reason, $R$ and $Y$ look
the same for both the on-shell and off-shell metrics. The parameters $m$ and $a$  enter the expressions for the
currents only via $R_H$ and $Y_A$.

The chiral current for the $U$-state is
\ba
\ts{J}_{U}= \frac{1}{ 2}(\ts{J}_+ +\ts{J}_-)\, ,
\ea
with $R_-=R_H$ and $R_+=0$.
For this state there is no incoming chirality flux from the  past infinity ${\cal J}^{-}$, and the current vanishes at $H_-$.   For these initial conditions, the principal chiral current contains an outgoing null flux at ${\cal J}^{+}$.  The chiral flux  at ${\cal J}^{+}$ is given by
\be\label{JU2}
\ts{J}_U  \sim -\frac{R_H}{ 2 r^2}\ts{l}_- \, .
\ee
For this state there also exists a chirality flux through the horizon $H_+$
\be
\ts{J}_U|_{H_+}=-\frac{R_H}{ 2(r_H^2+y^2)}\ts{l}_{+}\, .
\ee


\subsubsection{Special case of the Kerr-Newman black hole}

For the Kerr-Newman black hole, the function $R_H$ is
\be
R_H=\frac{4Q^2y r_H^2}{(r_H^2+y^2)^2} ,
\ee
where $r_H$ is the radius of the horizon.

Calculations for the total flux of chirality  give the following expression
\ba
\dot{L}=-\frac{4\pi Q^2 r_H}{a^2}\Big[\arctan\big(\frac{a}{r_H}\big)-\frac{a r_H}{a^2+r_H^2}\Big] ,
\ea
At small $a\ll r_H$ it becomes
\ba\label{dL1}
\dot{L}\approx -\frac{8\pi Q^2}{3r_H^2} a .
\ea
If the chiral anomaly was due to the Dirac fermions (\ref{FERM}), then the contribution of the electromagnetic chiral anomaly
would be obtained by multiplying the result  (\ref{dL1}) by a factor $e^2/(8\pi^2)$.

\subsection{Current and dual current}

Let us now discuss other solutions of the equation for the chiral anomaly that are generated by the electromagnetic field.
Let us denote
\be
J_{A}^{\mu}=2{}^*\!F^{\mu\nu}A_{\nu}\hh
J_{B}^{\mu}=2F^{\mu\nu}B_{\nu}\, .
\ee
It is easy to check that both currents describe the same chiral anomaly
\be
J_{A\, ;\mu}^{\mu}=J_{B\, ;\mu}^{\mu}=\mathcal{P}_e .
\ee
When expressed in terms of $U$ and $V$ functions  (\ref{UV}) they take the form
\ba
&J_{A}^{\mu}=\frac{2QrV}{\Sigma}\delta_r^{\mu}+\frac{2PyU}{\Sigma}\delta_y^{\mu},\\
&J_{B}^{\mu}=-\frac{2PrU}{\Sigma}\delta_r^{\mu}-\frac{2QyV}{\Sigma}\delta_y^{\mu} .
\ea

These currents are not gauge invariant. For the gauge transformation
\be
A_{\mu}\to \tilde{A}_{\mu}=A_{\mu}+\lambda_{;\mu}
\ee
the current $J_{A}^{\mu}$ gets a supplement
\be \n{GAUG}
\Delta J_{A}^{\mu}=2\,{}^*\!F^{\mu\nu}\lambda_{;\nu} \, .
\ee
Similarly, for \be
B_{\mu}\to \tilde{B}_{\mu}=B_{\mu}+\sigma_{;\mu}
\ee
the current $J_{B}^{\mu}$ gets a supplement
\be \n{GAUG}
\Delta {J}_B^{\mu}=2 F^{\mu\nu}\sigma_{;\nu} \, .
\ee
As a consequence of source-free Maxwell's equations both $\Delta {J}_A^{\mu}$ and $\Delta {J}_B^{\mu}$ are conserved for arbitrary gauge functions $\lambda$ and $\sigma$
\ba
&(\Delta{J}_A^{\mu})_{;\mu}=2\big({}^*\!F^{\mu\nu}{}_{;\mu}\lambda_{;\nu}+{}^*\!F^{\mu\nu}{}\lambda_{;\nu\mu}\big)=0,\\
&(\Delta{J}_B^{\mu})_{;\mu}=2\big(F^{\mu\nu}{}_{;\mu}\sigma_{;\nu}+F^{\mu\nu}{}\sigma_{;\nu\mu}\big)=0.
\ea

If the gauge functions $\lambda$ and $\sigma$ respect the spacetime symmetry, then they are functions of $r$ and $y$. It is easy to check that the field $F_{\mu\nu}$ for the potential (\ref{POTEN}) possesses the following property: It does not vanish only if one of its indices (say $\mu$) takes values in the $(r,y)$ sector, while the other (say $\nu$) takes values in the $(\tau,\psi)$ sector. This implies that the dual tensor $^*\!F^{\mu\nu}$ has the same property. Since the gradient of $\lambda$ has non-vanishing components $\lambda_{,r}$ and  $\lambda_{,y}$, the vector $\Delta  {J}_A^{\mu}$ has only $\tau$ and $\psi$ components. This property is valid for $\Delta  {J}_B^{\mu}$ as well. This means that the gauge transformations that respect the spacetime symmetry will leave the $r$ and $y$ components of the current invariant.

Denote
\be
J_s^{\mu}=\frac{1}{ 2}(J_{A}^{\mu}+J_{B}^{\mu})\hh
J_a^{\mu}=\frac{1}{ 2}(J_{A}^{\mu}-J_{B}^{\mu})\, .
\ee
Then one has
\be
J_{s\, ;\mu}^{\mu}=\mathcal{P}_e\hh  J_{a\, ;\mu}^{\mu}=0\, .
\ee

The currents $J_s^{\mu}$ and $J_a^{\mu}$ can be written in terms of the $R$- and $Y$-currents that we found in the previous subsection. Namely, one has
\ba
J_s^{\mu}=&\frac{1}{2 \Sigma}[R\delta_r^{\mu}+Y\delta_y^{\mu}] ,\\
J_a^{\mu}=&\frac{1}{2 \Sigma}[R\delta_r^{\mu}-Y\delta_y^{\mu}]
+\frac{2PU}{\Sigma}[r \delta_r^{\mu}+y\delta_y^{\mu}] ,
\ea
where functions $R$, $Y$, and $U$ were defined in (\ref{RY}) and (\ref{UV}), respectively.
Like before, one must require that at least one of the two charges $Q$ and $P$ vanishes in order to guarantee regularity at the symmetry axes.
The above chiral currents can be ``updated" to describe chirality fluxes at infinity for the $B$, $H$, and $U$ states.


\section{Discussion}
\label{s7}

In this paper, we discussed solutions to the chiral anomaly equation in a given spacetime metric.
At first, we considered a wide class of metrics which possess a principal Killing-Yano tensor.
We call such a  metric which contains two arbitrary functions of one variable $\Delta_r(r)$ and $\Delta_y(y)$
an ``off-shell" metric. Off-shell metrics contain two commuting Killing vector fields. This property allows one
to reduce the study of vector and tensor fields, as well as equations for the chiral anomaly, to the study of objects and
 equations in a special 2D space $S$. This procedure was developed by Geroch \cite{Geroch_2}, and we adapted his
 approach for our problem. In addition to Killing vectors, the off-shell metric also possesses a rank-two symmetric
 Killing tensor which is connected with a hidden symmetry of the metric. We obtained an expression for the
 chiral current, which is a solution of the chiral anomaly equation, and which respects both explicit and hidden
 symmetries of the off-shell metric. Next, we demonstrated that there exist special chiral anomaly currents
 that also respect the symmetries of the metric, which are directed along the principal null rays of the metric. We call such solutions
 ``principal chiral currents". These currents describe either incoming or outgoing polarization fluxes.
 We demonstrated that such currents can be chosen so that they satisfy regularity conditions both
  at the horizons and at the symmetry axes. We obtained contributions to the chiral anomaly currents
  generated by the Pontryagin invariant $\mathcal{P}=-\frac{1}{2}{}^*\!R^{\alpha\beta\mu\nu}R_{\alpha\beta\mu\nu}$,
  as well by the electromagnetic field invariant $\mathcal{P}_e={}^*\!F^{\alpha\beta}F_{\alpha\beta}$ in the
  case where such a field is present.

Next, we obtained expressions for the principal current for the special cases of Kerr and Kerr-Newman black holes.
For these spacetimes, if the principal chiral current initially vanishes at the past horizon and at the past null infinity,
then it has non-vanishing components at the future horizon and at the future null infinity. These components describe the
chirality flux into the black hole, and the chirality flux radiated by the black hole towards infinity, respectively.

Plot~\ref{Fig_7} shows the chiral flux  for the Kerr black hole, which is computed using the principal chiral current. Its angular asymmetry is correlated with the asymmetry of the Pontryagin invariant calculated at the horizon.

Using a general expression for the Chern-Simons nonconserved chiral current, we calculated its components for the special choice of the Darboux tetrad associated with the off-shell metric (\ref{ds2}), and used these results to find fluxes of the chirality at infinity and at the horizon. Plot~\ref{Fig_9} shows the chirality flux   at infinity as a function of the angle for the Kerr black hole. Comparing this plot with a similar plot ~\ref{Fig_7} for the flux of the principal current, one can see that  both currents have qualitatively the same behavior.
However, the Chern-Simons current is smaller, approximately by a factor between 3 and 10 depending on the rotation parameter. This is partly connected with the following: Both currents correctly reproduce the gravitational chiral anomaly, but the Chern-Simons current also contains a $y-$component, which is responsible for a part of the anomaly. For the principal current, this component is absent.

It is instructive to compare plots \ref{Fig_7} and \ref{Fig_9} with Fig.~3a of the paper \cite{Leahy:1979} for the emission rate of the number of neutrinos minus the number of antineutrinos,  which was calculated numerically. One can see a qualitative similarity between these plots. One can also conclude that the anomalous Chern-Simons current is numerically closer to the results of the paper \cite{Leahy:1979}.

In the papers \cite{Flachi:2017vlp,Stone:2018zel}, the authors discussed the axial current at finite rotation and temperature
in curved spacetime. They proposed an expression for the chiral current component along the axis of rotation of the system, which for the massless fermions and slow rotation takes the form
\be \n{JJTT}
J^z=\pm\Big(
\frac{T^2}{12} -\frac{R}{96\pi^2}\Big)\Omega ,
\ee
where $T$ is the temperature of the system, $\Omega$ is its angular velocity, and $R$ is the scalar curvature. For the Kerr black hole $R=0$, and one has
\be \n{JzJz}
J^z=\pm\frac{T^2}{12}\Omega\, .
\ee
As a rough approximate model, one may consider a rigidly rotating cylinder with the effective cross-section
$A\ins{eff}$ and angular velocity $\Omega_H\approx a/(4m^2)$. With the proper choice of sign, and after the substitution of the black hole temperature $T_H\approx 1/(8\pi m)$ into (\ref{JzJz}), we estimate the corresponding rate of chiral emission
\ba\label{dotLeff}
\dot{L}_\ins{eff}=- A\ins{eff} \frac{T_{H}^2}{12}\Omega_{H}.
\ea
In sections \ref{sec4C} and  \ref{sec5B}, we computed the total rate of chirality emission from the black hole, see relations (\ref{LLmm}) and (\ref{LLCS1}).

There exists a clear similarity between formulas (\ref{dotLeff}) and  (\ref{LLmm}),(\ref{LLCS1}). They are proportional to the angular velocity of the black hole with some dimensionless coefficient. Now we can estimate the effective area of the cross-section for the rotating cylinder in the \cite{Flachi:2017vlp,Stone:2018zel} approach in order to reproduce our results (\ref{LLmm}) and (\ref{LLCS1}). For slowly rotating black holes we get
$A_\ins{eff}\approx 3A_H$ in the case of the principal current, and $A_\ins{eff}\approx A_H$ in the case of the Chern-Simons current.

In the last section of the paper, the electromagnetic field contribution of the chiral anomaly current for massless charged Dirac particles was calculated, and expressions for the chirality fluxes in the case of the Kerr-Newman black hole were obtained.
It would be interesting to generalize our approach to the case of accelerating Kerr–Newman black holes in (A)dS spacetime. The Chern–Pontryagin invariant for this case was computed and studied in \cite{Kraniotis:2021qah}.

Certainly, the chiral anomaly equation has a larger variety of solutions. After one finds a current which correctly
reproduces the chiral anomaly, one still has the freedom to add to it a solution of the homogeneous equation
for a conserved current. We demonstrated that if such a current respects the symmetry generated by the Killing
vectors, then it contains three arbitrary functions of two variables, $r$ and $y$. This ambiguity is further reduced
after imposing the hidden symmetry constraint.

Let us emphasize that if the current is calculated on the basis of a microscopic theory as a quantum average for a quantum state in a spacetime with some symmetry generated by the Killing vectors, one can impose a corresponding symmetry condition on the quantum state. In such a case one can expect that the macroscopic averaged current would respect the spacetime explicit symmetry. It would be interesting to find restrictions (if any) on the choice of a state imposed by the existence of the hidden symmetries. At the moment, it is unclear how to formally implement this requirement for a quantum state. For this reason, we proceeded in a different way in this paper.

The existence of the principal Killing-Yano tensor in the off-shell geometry implies that it has a very special property. Namely, this metric belongs to the Petrov type-D class of metrics, and has two degenerate principal null directions which coincide with eigenvectors of the Killing tensor (for more details, see \cite{Frolov:2017kze} and references therein).  We demonstrated that there exists a solution of the chiral anomaly equation for which the corresponding chiral current is parallel to the principal null direction. Under this condition, the partial differential equation for the current reduces to an ordinary differential equation, and its solution can be found by simple integration of the gravitational and electromagnetic anomaly invariants along the principal null geodesics. One can interpret this result by saying that such a solution describes a free propagation of the created spinning massless particles along principal null geodesics without additional scattering by the gravitational field.


\acknowledgments

The authors thank A. Penin for useful discussions. This work was partly supported by the Natural Sciences and Engineering Research Council of Canada. The authors are also grateful to the Killam Trust for its financial support.


\begin{thebibliography}{50}%
\makeatletter
\providecommand \@ifxundefined [1]{%
 \@ifx{#1\undefined}
}%
\providecommand \@ifnum [1]{%
 \ifnum #1\expandafter \@firstoftwo
 \else \expandafter \@secondoftwo
 \fi
}%
\providecommand \@ifx [1]{%
 \ifx #1\expandafter \@firstoftwo
 \else \expandafter \@secondoftwo
 \fi
}%
\providecommand \natexlab [1]{#1}%
\providecommand \enquote  [1]{``#1''}%
\providecommand \bibnamefont  [1]{#1}%
\providecommand \bibfnamefont [1]{#1}%
\providecommand \citenamefont [1]{#1}%
\providecommand \href@noop [0]{\@secondoftwo}%
\providecommand \href [0]{\begingroup \@sanitize@url \@href}%
\providecommand \@href[1]{\@@startlink{#1}\@@href}%
\providecommand \@@href[1]{\endgroup#1\@@endlink}%
\providecommand \@sanitize@url [0]{\catcode `\\12\catcode `\$12\catcode
  `\&12\catcode `\#12\catcode `\^12\catcode `\_12\catcode `\%12\relax}%
\providecommand \@@startlink[1]{}%
\providecommand \@@endlink[0]{}%
\providecommand \url  [0]{\begingroup\@sanitize@url \@url }%
\providecommand \@url [1]{\endgroup\@href {#1}{\urlprefix }}%
\providecommand \urlprefix  [0]{URL }%
\providecommand \Eprint [0]{\href }%
\providecommand \doibase [0]{http://dx.doi.org/}%
\providecommand \selectlanguage [0]{\@gobble}%
\providecommand \bibinfo  [0]{\@secondoftwo}%
\providecommand \bibfield  [0]{\@secondoftwo}%
\providecommand \translation [1]{[#1]}%
\providecommand \BibitemOpen [0]{}%
\providecommand \bibitemStop [0]{}%
\providecommand \bibitemNoStop [0]{.\EOS\space}%
\providecommand \EOS [0]{\spacefactor3000\relax}%
\providecommand \BibitemShut  [1]{\csname bibitem#1\endcsname}%
\let\auto@bib@innerbib\@empty
\bibitem [{\citenamefont {Mashhoon}(2003)}]{Mashhoon:2003}%
  \BibitemOpen
  \bibfield  {author} {\bibinfo {author} {\bibfnamefont {B.}~\bibnamefont
  {Mashhoon}},\ }\bibfield  {title} {\enquote {\bibinfo {title}
  {Gravitoelectromagnetism: A brief review},}\ }\href {\doibase
  10.48550/arXiv.gr-qc/0311030} {\  (\bibinfo {year} {2003}),\
  10.48550/arXiv.gr-qc/0311030},\ \Eprint {http://arxiv.org/abs/gr-qc/0311030}
  {arXiv:gr-qc/0311030} \BibitemShut {NoStop}%
\bibitem [{\citenamefont {Mathisson}(1937)}]{Mathisson:1937}%
  \BibitemOpen
  \bibfield  {author} {\bibinfo {author} {\bibfnamefont {M.}~\bibnamefont
  {Mathisson}},\ }\bibfield  {title} {\enquote {\bibinfo {title} {{Neue
  mechanik materieller systemes}},}\ }\href@noop {} {\bibfield  {journal}
  {\bibinfo  {journal} {Acta Phys. Polon.}\ }\textbf {\bibinfo {volume} {6}},\
  \bibinfo {pages} {163--2900} (\bibinfo {year} {1937})}\BibitemShut {NoStop}%
\bibitem [{\citenamefont {Papapetrou}(1951)}]{Papapetrou:1951pa}%
  \BibitemOpen
  \bibfield  {author} {\bibinfo {author} {\bibfnamefont {A.}~\bibnamefont
  {Papapetrou}},\ }\bibfield  {title} {\enquote {\bibinfo {title} {{Spinning
  test particles in general relativity. 1.}}}\ }\href {\doibase
  10.1098/rspa.1951.0200} {\bibfield  {journal} {\bibinfo  {journal} {Proc.
  Roy. Soc. Lond. A}\ }\textbf {\bibinfo {volume} {209}},\ \bibinfo {pages}
  {248--258} (\bibinfo {year} {1951})}\BibitemShut {NoStop}%
\bibitem [{\citenamefont {Dixon}(1970)}]{Dixon}%
  \BibitemOpen
  \bibfield  {author} {\bibinfo {author} {\bibfnamefont {W.~G.}\ \bibnamefont
  {Dixon}},\ }\bibfield  {title} {\enquote {\bibinfo {title} {{Dynamics of
  extended bodies in general relativity. I. Momentum and angular momentum}},}\
  }\href {\doibase 10.1098/rspa.1970.0020} {\bibfield  {journal} {\bibinfo
  {journal} {Proc. Roy. Soc. Lond. A}\ }\textbf {\bibinfo {volume} {314}},\
  \bibinfo {pages} {499--527} (\bibinfo {year} {1970})}\BibitemShut {NoStop}%
\bibitem [{\citenamefont {Wald}(1972)}]{WALD:1972}%
  \BibitemOpen
  \bibfield  {author} {\bibinfo {author} {\bibfnamefont {R.}~\bibnamefont
  {Wald}},\ }\bibfield  {title} {\enquote {\bibinfo {title} {Gravitational spin
  interaction},}\ }\href {\doibase 10.1103/PhysRevD.6.406} {\bibfield
  {journal} {\bibinfo  {journal} {Phys. Rev. D}\ }\textbf {\bibinfo {volume}
  {6}},\ \bibinfo {pages} {406--413} (\bibinfo {year} {1972})}\BibitemShut
  {NoStop}%
\bibitem [{\citenamefont {Leahy}\ and\ \citenamefont
  {Unruh}(1979)}]{Leahy:1979}%
  \BibitemOpen
  \bibfield  {author} {\bibinfo {author} {\bibfnamefont {D.~A.}\ \bibnamefont
  {Leahy}}\ and\ \bibinfo {author} {\bibfnamefont {W.~G.}\ \bibnamefont
  {Unruh}},\ }\bibfield  {title} {\enquote {\bibinfo {title} {{Angular
  dependence of neutrino emission from rotating black holes}},}\ }\href
  {\doibase 10.1103/PhysRevD.19.3509} {\bibfield  {journal} {\bibinfo
  {journal} {Phys. Rev. D}\ }\textbf {\bibinfo {volume} {19}},\ \bibinfo
  {pages} {3509--3515} (\bibinfo {year} {1979})}\BibitemShut {NoStop}%
\bibitem [{\citenamefont {Vilenkin}(1979)}]{Vilenkin:1979}%
  \BibitemOpen
  \bibfield  {author} {\bibinfo {author} {\bibfnamefont {A.}~\bibnamefont
  {Vilenkin}},\ }\bibfield  {title} {\enquote {\bibinfo {title} {{Macroscopic
  parity-violating effects: Neutrino fluxes from rotating black holes and in
  rotating thermal radiation}},}\ }\href {\doibase 10.1103/PhysRevD.20.1807}
  {\bibfield  {journal} {\bibinfo  {journal} {Phys. Rev. D}\ }\textbf {\bibinfo
  {volume} {20}},\ \bibinfo {pages} {1807--1812} (\bibinfo {year}
  {1979})}\BibitemShut {NoStop}%
\bibitem [{\citenamefont {Bolashenko}\ and\ \citenamefont
  {Frolov}(1989{\natexlab{a}})}]{Bolashenko:1989}%
  \BibitemOpen
  \bibfield  {author} {\bibinfo {author} {\bibfnamefont {P.~A.}\ \bibnamefont
  {Bolashenko}}\ and\ \bibinfo {author} {\bibfnamefont {V.~P.}\ \bibnamefont
  {Frolov}},\ }\bibfield  {title} {\enquote {\bibinfo {title} {{Density matrix
  and generating functional for quantized neutrino field in the space-time of a
  rotating black hole}},}\ }\href {\doibase doi.org/10.1007/BF01016914}
  {\bibfield  {journal} {\bibinfo  {journal} {Theoret. and Math. Phys.}\
  }\textbf {\bibinfo {volume} {78}},\ \bibinfo {pages} {31--41} (\bibinfo
  {year} {1989}{\natexlab{a}})}\BibitemShut {NoStop}%
\bibitem [{\citenamefont {Casals}\ \emph {et~al.}(2013)\citenamefont {Casals},
  \citenamefont {Dolan}, \citenamefont {Nolan}, \citenamefont {Ottewill},\ and\
  \citenamefont {Winstanley}}]{Casals:2012es}%
  \BibitemOpen
  \bibfield  {author} {\bibinfo {author} {\bibfnamefont {M.}~\bibnamefont
  {Casals}}, \bibinfo {author} {\bibfnamefont {S.~R.}\ \bibnamefont {Dolan}},
  \bibinfo {author} {\bibfnamefont {B.~C.}\ \bibnamefont {Nolan}}, \bibinfo
  {author} {\bibfnamefont {A.~C.}\ \bibnamefont {Ottewill}}, \ and\ \bibinfo
  {author} {\bibfnamefont {E.}~\bibnamefont {Winstanley}},\ }\bibfield  {title}
  {\enquote {\bibinfo {title} {{Quantization of fermions on Kerr
  space-time}},}\ }\href {\doibase 10.1103/PhysRevD.87.064027} {\bibfield
  {journal} {\bibinfo  {journal} {Phys. Rev. D}\ }\textbf {\bibinfo {volume}
  {87}},\ \bibinfo {pages} {064027} (\bibinfo {year} {2013})},\ \Eprint
  {http://arxiv.org/abs/1207.7089} {arXiv:1207.7089 [gr-qc]} \BibitemShut
  {NoStop}%
\bibitem [{\citenamefont {Bolashenko}\ and\ \citenamefont
  {Frolov}(1989{\natexlab{b}})}]{Bolashenko:1989trudy}%
  \BibitemOpen
  \bibfield  {author} {\bibinfo {author} {\bibfnamefont {P.~A.}\ \bibnamefont
  {Bolashenko}}\ and\ \bibinfo {author} {\bibfnamefont {V.~P.}\ \bibnamefont
  {Frolov}},\ }\bibfield  {title} {\enquote {\bibinfo {title} {{Quantum effects
  for massless spin particles in gravitational field of rotating black
  hole}},}\ }\href@noop {} {\bibfield  {journal} {\bibinfo  {journal} {Trudy
  Fizicheskogo Instituta imeni PN Levedeba, Akademiya Nauk SSSR (In Russian)}\
  }\textbf {\bibinfo {volume} {197}},\ \bibinfo {pages} {88--149} (\bibinfo
  {year} {1989}{\natexlab{b}})}\BibitemShut {NoStop}%
\bibitem [{\citenamefont {Casals}\ \emph {et~al.}(2009)\citenamefont {Casals},
  \citenamefont {Dolan}, \citenamefont {Kanti},\ and\ \citenamefont
  {Winstanley}}]{Casals_2009}%
  \BibitemOpen
  \bibfield  {author} {\bibinfo {author} {\bibfnamefont {M.}~\bibnamefont
  {Casals}}, \bibinfo {author} {\bibfnamefont {S.~R.}\ \bibnamefont {Dolan}},
  \bibinfo {author} {\bibfnamefont {P.}~\bibnamefont {Kanti}}, \ and\ \bibinfo
  {author} {\bibfnamefont {E.}~\bibnamefont {Winstanley}},\ }\bibfield  {title}
  {\enquote {\bibinfo {title} {Angular profile of emission of non-zero spin
  fields from a higher-dimensional black hole},}\ }\href {\doibase
  10.1016/j.physletb.2009.08.062} {\bibfield  {journal} {\bibinfo  {journal}
  {Physics Letters B}\ }\textbf {\bibinfo {volume} {680}},\ \bibinfo {pages}
  {365--370} (\bibinfo {year} {2009})}\BibitemShut {NoStop}%
\bibitem [{\citenamefont {Dolgov}\ \emph {et~al.}(1988)\citenamefont {Dolgov},
  \citenamefont {Khriplovich},\ and\ \citenamefont {Zakharov}}]{DOLGOV:1988}%
  \BibitemOpen
  \bibfield  {author} {\bibinfo {author} {\bibfnamefont {A.~D.}\ \bibnamefont
  {Dolgov}}, \bibinfo {author} {\bibfnamefont {I.~B.}\ \bibnamefont
  {Khriplovich}}, \ and\ \bibinfo {author} {\bibfnamefont {V.~I.}\ \bibnamefont
  {Zakharov}},\ }\bibfield  {title} {\enquote {\bibinfo {title} {Macroscopic
  manifestations of the chiral anomaly in a gravitational field},}\ }\href
  {\doibase https://doi.org/10.1016/0550-3213(88)90460-9} {\bibfield  {journal}
  {\bibinfo  {journal} {Nuclear Physics B}\ }\textbf {\bibinfo {volume}
  {309}},\ \bibinfo {pages} {591--596} (\bibinfo {year} {1988})}\BibitemShut
  {NoStop}%
\bibitem [{\citenamefont {Dolgov}\ \emph
  {et~al.}(1989{\natexlab{a}})\citenamefont {Dolgov}, \citenamefont
  {Khriplovich}, \citenamefont {Vainshtein},\ and\ \citenamefont
  {Zakharov}}]{Dolgov:1989a}%
  \BibitemOpen
  \bibfield  {author} {\bibinfo {author} {\bibfnamefont {A.~D.}\ \bibnamefont
  {Dolgov}}, \bibinfo {author} {\bibfnamefont {I.~B.}\ \bibnamefont
  {Khriplovich}}, \bibinfo {author} {\bibfnamefont {A.~I.}\ \bibnamefont
  {Vainshtein}}, \ and\ \bibinfo {author} {\bibfnamefont {V.~I.}\ \bibnamefont
  {Zakharov}},\ }\bibfield  {title} {\enquote {\bibinfo {title} {{Photonic
  Chiral Current and Its Anomaly in a Gravitational Field}},}\ }\href {\doibase
  10.1016/0550-3213(89)90451-3} {\bibfield  {journal} {\bibinfo  {journal}
  {Nucl. Phys. B}\ }\textbf {\bibinfo {volume} {315}},\ \bibinfo {pages}
  {138--152} (\bibinfo {year} {1989}{\natexlab{a}})}\BibitemShut {NoStop}%
\bibitem [{\citenamefont {Dolgov}\ \emph
  {et~al.}(1989{\natexlab{b}})\citenamefont {Dolgov}, \citenamefont
  {Khriplovich}, \citenamefont {Vainshtein},\ and\ \citenamefont
  {Zakharov}}]{DOLGOV:1989b}%
  \BibitemOpen
  \bibfield  {author} {\bibinfo {author} {\bibfnamefont {A.~D.}\ \bibnamefont
  {Dolgov}}, \bibinfo {author} {\bibfnamefont {I.~B.}\ \bibnamefont
  {Khriplovich}}, \bibinfo {author} {\bibfnamefont {A.~I.}\ \bibnamefont
  {Vainshtein}}, \ and\ \bibinfo {author} {\bibfnamefont {V.~I.}\ \bibnamefont
  {Zakharov}},\ }\bibfield  {title} {\enquote {\bibinfo {title} {Vanishing of
  the chiral anomaly for an antisymmetric tensor field},}\ }\href {\doibase
  https://doi.org/10.1016/0550-3213(89)90513-0} {\bibfield  {journal} {\bibinfo
   {journal} {Nuclear Physics B}\ }\textbf {\bibinfo {volume} {313}},\ \bibinfo
  {pages} {73--79} (\bibinfo {year} {1989}{\natexlab{b}})}\BibitemShut
  {NoStop}%
\bibitem [{\citenamefont {Bertlmann}(2000)}]{Bertlmann:2000}%
  \BibitemOpen
  \bibfield  {author} {\bibinfo {author} {\bibfnamefont {R.~A.}\ \bibnamefont
  {Bertlmann}},\ }\href {\doibase 10.1093/acprof:oso/9780198507628.001.0001}
  {\emph {\bibinfo {title} {Anomalies in quantum field theory}}},\
  Vol.~\bibinfo {volume} {91}\ (\bibinfo  {publisher} {Oxford university
  press},\ \bibinfo {year} {2000})\BibitemShut {NoStop}%
\bibitem [{\citenamefont {Fujikawa}\ and\ \citenamefont
  {Suzuki}(2004)}]{Fujikawa2004path}%
  \BibitemOpen
  \bibfield  {author} {\bibinfo {author} {\bibfnamefont {K.}~\bibnamefont
  {Fujikawa}}\ and\ \bibinfo {author} {\bibfnamefont {H.}~\bibnamefont
  {Suzuki}},\ }\href {\doibase 10.1093/acprof:oso/9780198529132.001.0001}
  {\emph {\bibinfo {title} {Path integrals and quantum anomalies}}},\ \bibinfo
  {number} {122}\ (\bibinfo  {publisher} {Oxford University Press},\ \bibinfo
  {year} {2004})\BibitemShut {NoStop}%
\bibitem [{\citenamefont {Zakharov}(2013)}]{Zakharov:2012vv}%
  \BibitemOpen
  \bibfield  {author} {\bibinfo {author} {\bibfnamefont {V.~I.}\ \bibnamefont
  {Zakharov}},\ }\bibfield  {title} {\enquote {\bibinfo {title} {{Chiral
  Magnetic Effect in Hydrodynamic Approximation}},}\ }\href {\doibase
  10.1007/978-3-642-37305-3_11} {\bibfield  {journal} {\bibinfo  {journal}
  {Lect. Notes Phys.}\ }\textbf {\bibinfo {volume} {871}},\ \bibinfo {pages}
  {295--330} (\bibinfo {year} {2013})},\ \Eprint
  {http://arxiv.org/abs/1210.2186} {arXiv:1210.2186 [hep-ph]} \BibitemShut
  {NoStop}%
\bibitem [{\citenamefont {Chernodub}\ \emph {et~al.}(2022)\citenamefont
  {Chernodub}, \citenamefont {Ferreiros}, \citenamefont {Grushin},
  \citenamefont {Landsteiner},\ and\ \citenamefont
  {Vozmediano}}]{Chernodub:2021nff}%
  \BibitemOpen
  \bibfield  {author} {\bibinfo {author} {\bibfnamefont {Maxim~N.}\
  \bibnamefont {Chernodub}}, \bibinfo {author} {\bibfnamefont {Yago}\
  \bibnamefont {Ferreiros}}, \bibinfo {author} {\bibfnamefont {Adolfo~G.}\
  \bibnamefont {Grushin}}, \bibinfo {author} {\bibfnamefont {Karl}\
  \bibnamefont {Landsteiner}}, \ and\ \bibinfo {author} {\bibfnamefont
  {Mar\'\i{}a A.~H.}\ \bibnamefont {Vozmediano}},\ }\bibfield  {title}
  {\enquote {\bibinfo {title} {{Thermal transport, geometry, and anomalies}},}\
  }\href {\doibase 10.1016/j.physrep.2022.06.002} {\bibfield  {journal}
  {\bibinfo  {journal} {Phys. Rept.}\ }\textbf {\bibinfo {volume} {977}},\
  \bibinfo {pages} {1--58} (\bibinfo {year} {2022})},\ \Eprint
  {http://arxiv.org/abs/2110.05471} {arXiv:2110.05471 [cond-mat.mes-hall]}
  \BibitemShut {NoStop}%
\bibitem [{\citenamefont {Kimura}(1969)}]{Kimura:1969}%
  \BibitemOpen
  \bibfield  {author} {\bibinfo {author} {\bibfnamefont {T.}~\bibnamefont
  {Kimura}},\ }\bibfield  {title} {\enquote {\bibinfo {title} {{Divergence of
  Axial-Vector Current in the Gravitational Field}},}\ }\href {\doibase
  10.1143/PTP.42.1191} {\bibfield  {journal} {\bibinfo  {journal} {Progress of
  Theoretical Physics}\ }\textbf {\bibinfo {volume} {42}},\ \bibinfo {pages}
  {1191--1205} (\bibinfo {year} {1969})}\BibitemShut {NoStop}%
\bibitem [{\citenamefont {Delbourgo}\ and\ \citenamefont
  {Salam}(1972)}]{SALAM}%
  \BibitemOpen
  \bibfield  {author} {\bibinfo {author} {\bibfnamefont {R.}~\bibnamefont
  {Delbourgo}}\ and\ \bibinfo {author} {\bibfnamefont {A.}~\bibnamefont
  {Salam}},\ }\bibfield  {title} {\enquote {\bibinfo {title} {The gravitational
  correction to pcac},}\ }\href {\doibase
  https://doi.org/10.1016/0370-2693(72)90825-8} {\bibfield  {journal} {\bibinfo
   {journal} {Physics Letters B}\ }\textbf {\bibinfo {volume} {40}},\ \bibinfo
  {pages} {381--382} (\bibinfo {year} {1972})}\BibitemShut {NoStop}%
\bibitem [{\citenamefont {Eguchi}\ and\ \citenamefont {Freund}(1976)}]{EGUCHI}%
  \BibitemOpen
  \bibfield  {author} {\bibinfo {author} {\bibfnamefont {T.}~\bibnamefont
  {Eguchi}}\ and\ \bibinfo {author} {\bibfnamefont {P.~G.~O.}\ \bibnamefont
  {Freund}},\ }\bibfield  {title} {\enquote {\bibinfo {title} {Quantum gravity
  and world topology},}\ }\href {\doibase 10.1103/PhysRevLett.37.1251}
  {\bibfield  {journal} {\bibinfo  {journal} {Phys. Rev. Lett.}\ }\textbf
  {\bibinfo {volume} {37}},\ \bibinfo {pages} {1251--1254} (\bibinfo {year}
  {1976})}\BibitemShut {NoStop}%
\bibitem [{\citenamefont {Prokhorov}\ \emph {et~al.}(2022)\citenamefont
  {Prokhorov}, \citenamefont {Teryaev},\ and\ \citenamefont
  {Zakharov}}]{Prokhorov:2022rna}%
  \BibitemOpen
  \bibfield  {author} {\bibinfo {author} {\bibfnamefont {G.~Yu.}\ \bibnamefont
  {Prokhorov}}, \bibinfo {author} {\bibfnamefont {O.~V.}\ \bibnamefont
  {Teryaev}}, \ and\ \bibinfo {author} {\bibfnamefont {V.~I.}\ \bibnamefont
  {Zakharov}},\ }\bibfield  {title} {\enquote {\bibinfo {title} {{Gravitational
  chiral anomaly for spin 3/2 field interacting with spin 1/2 field}},}\ }\href
  {\doibase 10.1103/PhysRevD.106.025022} {\bibfield  {journal} {\bibinfo
  {journal} {Phys. Rev. D}\ }\textbf {\bibinfo {volume} {106}},\ \bibinfo
  {pages} {025022} (\bibinfo {year} {2022})},\ \Eprint
  {http://arxiv.org/abs/2202.02168} {arXiv:2202.02168 [hep-th]} \BibitemShut
  {NoStop}%
\bibitem [{\citenamefont {Misner}\ \emph {et~al.}(1973)\citenamefont {Misner},
  \citenamefont {Thorne},\ and\ \citenamefont {Wheeler}}]{Misner:1973prb}%
  \BibitemOpen
  \bibfield  {author} {\bibinfo {author} {\bibfnamefont {C.~W.}\ \bibnamefont
  {Misner}}, \bibinfo {author} {\bibfnamefont {K.~S.}\ \bibnamefont {Thorne}},
  \ and\ \bibinfo {author} {\bibfnamefont {J.~A.}\ \bibnamefont {Wheeler}},\
  }\href@noop {} {\emph {\bibinfo {title} {{Gravitation}}}}\ (\bibinfo
  {publisher} {W. H. Freeman},\ \bibinfo {address} {San Francisco},\ \bibinfo
  {year} {1973})\BibitemShut {NoStop}%
\bibitem [{\citenamefont {Tong}()}]{CAMBRIDGE}%
  \BibitemOpen
  \bibfield  {author} {\bibinfo {author} {\bibfnamefont {D.}~\bibnamefont
  {Tong}},\ }\href {https://www.damtp.cam.ac.uk/user/tong/gaugetheory.html}
  {\emph {\bibinfo {title} {{Lectures on Gauge Theory}}}}\BibitemShut {NoStop}%
\bibitem [{\citenamefont {Dolgov}\ \emph {et~al.}(1987)\citenamefont {Dolgov},
  \citenamefont {Khriplovich},\ and\ \citenamefont {Zakharov}}]{Dolgov:1987yp}%
  \BibitemOpen
  \bibfield  {author} {\bibinfo {author} {\bibfnamefont {A.~D.}\ \bibnamefont
  {Dolgov}}, \bibinfo {author} {\bibfnamefont {I.~B.}\ \bibnamefont
  {Khriplovich}}, \ and\ \bibinfo {author} {\bibfnamefont {V.~I.}\ \bibnamefont
  {Zakharov}},\ }\bibfield  {title} {\enquote {\bibinfo {title} {{Chiral Boson
  Anomaly in a Gravitational Field}},}\ }\href@noop {} {\bibfield  {journal}
  {\bibinfo  {journal} {JETP Lett.}\ }\textbf {\bibinfo {volume} {45}},\
  \bibinfo {pages} {651--653} (\bibinfo {year} {1987})}\BibitemShut {NoStop}%
\bibitem [{\citenamefont {Agullo}\ \emph {et~al.}(2018)\citenamefont {Agullo},
  \citenamefont {del Rio},\ and\ \citenamefont
  {Navarro-Salas}}]{Agullo:2018nfv}%
  \BibitemOpen
  \bibfield  {author} {\bibinfo {author} {\bibfnamefont {I.}~\bibnamefont
  {Agullo}}, \bibinfo {author} {\bibfnamefont {A.}~\bibnamefont {del Rio}}, \
  and\ \bibinfo {author} {\bibfnamefont {J.}~\bibnamefont {Navarro-Salas}},\
  }\bibfield  {title} {\enquote {\bibinfo {title} {{Classical and quantum
  aspects of electric-magnetic duality rotations in curved spacetimes}},}\
  }\href {\doibase 10.1103/PhysRevD.98.125001} {\bibfield  {journal} {\bibinfo
  {journal} {Phys. Rev. D}\ }\textbf {\bibinfo {volume} {98}},\ \bibinfo
  {pages} {125001} (\bibinfo {year} {2018})},\ \Eprint
  {http://arxiv.org/abs/1810.08085} {arXiv:1810.08085 [gr-qc]} \BibitemShut
  {NoStop}%
\bibitem [{\citenamefont {Galaverni}\ and\ \citenamefont
  {Gionti}(2021)}]{Galaverni:2021}%
  \BibitemOpen
  \bibfield  {author} {\bibinfo {author} {\bibfnamefont {M.}~\bibnamefont
  {Galaverni}}\ and\ \bibinfo {author} {\bibfnamefont {S.~J.~G.}\ \bibnamefont
  {Gionti}},\ }\bibfield  {title} {\enquote {\bibinfo {title} {Photon helicity
  and quantum anomalies in curved spacetimes},}\ }\href {\doibase
  10.1007/s10714-021-02817-z} {\bibfield  {journal} {\bibinfo  {journal}
  {General Relativity and Gravitation}\ }\textbf {\bibinfo {volume} {53}}
  (\bibinfo {year} {2021}),\ 10.1007/s10714-021-02817-z}\BibitemShut {NoStop}%
\bibitem [{\citenamefont {Bastianelli}\ and\ \citenamefont {van
  Nieuwenhuizen}(2006)}]{nieu:2006}%
  \BibitemOpen
  \bibfield  {author} {\bibinfo {author} {\bibfnamefont {F.}~\bibnamefont
  {Bastianelli}}\ and\ \bibinfo {author} {\bibfnamefont {P.}~\bibnamefont {van
  Nieuwenhuizen}},\ }\href {\doibase 10.1017/CBO9780511535031} {\emph {\bibinfo
  {title} {Path Integrals and Anomalies in Curved Space}}},\ Cambridge
  Monographs on Mathematical Physics\ (\bibinfo  {publisher} {Cambridge
  University Press},\ \bibinfo {year} {2006})\BibitemShut {NoStop}%
\bibitem [{\citenamefont {Christensen}\ and\ \citenamefont
  {Fulling}(1977)}]{Christensen:1977jc}%
  \BibitemOpen
  \bibfield  {author} {\bibinfo {author} {\bibfnamefont {S.~M.}\ \bibnamefont
  {Christensen}}\ and\ \bibinfo {author} {\bibfnamefont {S.~A.}\ \bibnamefont
  {Fulling}},\ }\bibfield  {title} {\enquote {\bibinfo {title} {{Trace
  Anomalies and the Hawking Effect}},}\ }\href {\doibase
  10.1103/PhysRevD.15.2088} {\bibfield  {journal} {\bibinfo  {journal} {Phys.
  Rev. D}\ }\textbf {\bibinfo {volume} {15}},\ \bibinfo {pages} {2088--2104}
  (\bibinfo {year} {1977})}\BibitemShut {NoStop}%
\bibitem [{\citenamefont {Robinson}\ and\ \citenamefont
  {Wilczek}(2005)}]{Robinson:2005pd}%
  \BibitemOpen
  \bibfield  {author} {\bibinfo {author} {\bibfnamefont {S.~P.}\ \bibnamefont
  {Robinson}}\ and\ \bibinfo {author} {\bibfnamefont {F.}~\bibnamefont
  {Wilczek}},\ }\bibfield  {title} {\enquote {\bibinfo {title} {{A Relationship
  between Hawking radiation and gravitational anomalies}},}\ }\href {\doibase
  10.1103/PhysRevLett.95.011303} {\bibfield  {journal} {\bibinfo  {journal}
  {Phys. Rev. Lett.}\ }\textbf {\bibinfo {volume} {95}},\ \bibinfo {pages}
  {011303} (\bibinfo {year} {2005})},\ \Eprint
  {http://arxiv.org/abs/gr-qc/0502074} {arXiv:gr-qc/0502074} \BibitemShut
  {NoStop}%
\bibitem [{\citenamefont {Iso}\ \emph {et~al.}(2006{\natexlab{a}})\citenamefont
  {Iso}, \citenamefont {Umetsu},\ and\ \citenamefont {Wilczek}}]{Iso:2006wa}%
  \BibitemOpen
  \bibfield  {author} {\bibinfo {author} {\bibfnamefont {S.}~\bibnamefont
  {Iso}}, \bibinfo {author} {\bibfnamefont {H.}~\bibnamefont {Umetsu}}, \ and\
  \bibinfo {author} {\bibfnamefont {F.}~\bibnamefont {Wilczek}},\ }\bibfield
  {title} {\enquote {\bibinfo {title} {{Hawking radiation from charged black
  holes via gauge and gravitational anomalies}},}\ }\href {\doibase
  10.1103/PhysRevLett.96.151302} {\bibfield  {journal} {\bibinfo  {journal}
  {Phys. Rev. Lett.}\ }\textbf {\bibinfo {volume} {96}},\ \bibinfo {pages}
  {151302} (\bibinfo {year} {2006}{\natexlab{a}})},\ \Eprint
  {http://arxiv.org/abs/hep-th/0602146} {arXiv:hep-th/0602146} \BibitemShut
  {NoStop}%
\bibitem [{\citenamefont {Iso}\ \emph {et~al.}(2006{\natexlab{b}})\citenamefont
  {Iso}, \citenamefont {Umetsu},\ and\ \citenamefont {Wilczek}}]{Iso:2006ut}%
  \BibitemOpen
  \bibfield  {author} {\bibinfo {author} {\bibfnamefont {S.}~\bibnamefont
  {Iso}}, \bibinfo {author} {\bibfnamefont {H.}~\bibnamefont {Umetsu}}, \ and\
  \bibinfo {author} {\bibfnamefont {F.}~\bibnamefont {Wilczek}},\ }\bibfield
  {title} {\enquote {\bibinfo {title} {{Anomalies, Hawking radiations and
  regularity in rotating black holes}},}\ }\href {\doibase
  10.1103/PhysRevD.74.044017} {\bibfield  {journal} {\bibinfo  {journal} {Phys.
  Rev. D}\ }\textbf {\bibinfo {volume} {74}},\ \bibinfo {pages} {044017}
  (\bibinfo {year} {2006}{\natexlab{b}})},\ \Eprint
  {http://arxiv.org/abs/hep-th/0606018} {arXiv:hep-th/0606018} \BibitemShut
  {NoStop}%
\bibitem [{\citenamefont {Banerjee}\ and\ \citenamefont
  {Kulkarni}(2008)}]{Banerjee:2007qs}%
  \BibitemOpen
  \bibfield  {author} {\bibinfo {author} {\bibfnamefont {R.}~\bibnamefont
  {Banerjee}}\ and\ \bibinfo {author} {\bibfnamefont {S.}~\bibnamefont
  {Kulkarni}},\ }\bibfield  {title} {\enquote {\bibinfo {title} {{Hawking
  radiation and covariant anomalies}},}\ }\href {\doibase
  10.1103/PhysRevD.77.024018} {\bibfield  {journal} {\bibinfo  {journal} {Phys.
  Rev. D}\ }\textbf {\bibinfo {volume} {77}},\ \bibinfo {pages} {024018}
  (\bibinfo {year} {2008})},\ \Eprint {http://arxiv.org/abs/0707.2449}
  {arXiv:0707.2449 [hep-th]} \BibitemShut {NoStop}%
\bibitem [{\citenamefont {Blaer}\ \emph {et~al.}(1981)\citenamefont {Blaer},
  \citenamefont {Christ},\ and\ \citenamefont {Tang}}]{Blaer:1981ps}%
  \BibitemOpen
  \bibfield  {author} {\bibinfo {author} {\bibfnamefont {A.~S.}\ \bibnamefont
  {Blaer}}, \bibinfo {author} {\bibfnamefont {N.~H.}\ \bibnamefont {Christ}}, \
  and\ \bibinfo {author} {\bibfnamefont {Ju-Fei}\ \bibnamefont {Tang}},\
  }\bibfield  {title} {\enquote {\bibinfo {title} {{Anomalous fermion
  production by a Julia-Zee dyon}},}\ }\href {\doibase
  10.1103/PhysRevLett.47.1364} {\bibfield  {journal} {\bibinfo  {journal}
  {Phys. Rev. Lett.}\ }\textbf {\bibinfo {volume} {47}},\ \bibinfo {pages}
  {1364} (\bibinfo {year} {1981})}\BibitemShut {NoStop}%
\bibitem [{\citenamefont {Frolov}\ \emph {et~al.}(2017)\citenamefont {Frolov},
  \citenamefont {Krtous},\ and\ \citenamefont {Kubiznak}}]{Frolov:2017kze}%
  \BibitemOpen
  \bibfield  {author} {\bibinfo {author} {\bibfnamefont {V.~P.}\ \bibnamefont
  {Frolov}}, \bibinfo {author} {\bibfnamefont {P.}~\bibnamefont {Krtous}}, \
  and\ \bibinfo {author} {\bibfnamefont {D.}~\bibnamefont {Kubiznak}},\
  }\bibfield  {title} {\enquote {\bibinfo {title} {{Black holes, hidden
  symmetries, and complete integrability}},}\ }\href {\doibase
  10.1007/s41114-017-0009-9} {\bibfield  {journal} {\bibinfo  {journal} {Living
  Rev. Rel.}\ }\textbf {\bibinfo {volume} {20}},\ \bibinfo {pages} {6}
  (\bibinfo {year} {2017})},\ \Eprint {http://arxiv.org/abs/1705.05482}
  {arXiv:1705.05482 [gr-qc]} \BibitemShut {NoStop}%
\bibitem [{\citenamefont {Frolov}\ and\ \citenamefont
  {Zelnikov}(2011)}]{FrolovZelnikov:2011}%
  \BibitemOpen
  \bibfield  {author} {\bibinfo {author} {\bibfnamefont {V.~P.}\ \bibnamefont
  {Frolov}}\ and\ \bibinfo {author} {\bibfnamefont {A.}~\bibnamefont
  {Zelnikov}},\ }\href {http://books.google.ca/books?id=UTGFNAEACAAJ} {\emph
  {\bibinfo {title} {{Introduction to Black Hole Physics}}}}\ (\bibinfo
  {publisher} {Oxford University Press},\ \bibinfo {year} {2011})\BibitemShut
  {NoStop}%
\bibitem [{\citenamefont {Geroch}(1971)}]{Geroch_1}%
  \BibitemOpen
  \bibfield  {author} {\bibinfo {author} {\bibfnamefont {R.~P.}\ \bibnamefont
  {Geroch}},\ }\bibfield  {title} {\enquote {\bibinfo {title} {{A Method for
  generating solutions of Einstein's equations}},}\ }\href {\doibase
  10.1063/1.1665681} {\bibfield  {journal} {\bibinfo  {journal} {J. Math.
  Phys.}\ }\textbf {\bibinfo {volume} {12}},\ \bibinfo {pages} {918--924}
  (\bibinfo {year} {1971})}\BibitemShut {NoStop}%
\bibitem [{\citenamefont {Geroch}(1972)}]{Geroch_2}%
  \BibitemOpen
  \bibfield  {author} {\bibinfo {author} {\bibfnamefont {R.~P.}\ \bibnamefont
  {Geroch}},\ }\bibfield  {title} {\enquote {\bibinfo {title} {{A Method for
  generating new solutions of Einstein's equation. 2}},}\ }\href {\doibase
  10.1063/1.1665990} {\bibfield  {journal} {\bibinfo  {journal} {J. Math.
  Phys.}\ }\textbf {\bibinfo {volume} {13}},\ \bibinfo {pages} {394--404}
  (\bibinfo {year} {1972})}\BibitemShut {NoStop}%
\bibitem [{\citenamefont {Stephani}\ \emph {et~al.}(2003)\citenamefont
  {Stephani}, \citenamefont {Kramer}, \citenamefont {MacCallum}, \citenamefont
  {Hoenselaers},\ and\ \citenamefont {Herlt}}]{kramer:2003}%
  \BibitemOpen
  \bibfield  {author} {\bibinfo {author} {\bibfnamefont {H.}~\bibnamefont
  {Stephani}}, \bibinfo {author} {\bibfnamefont {D.}~\bibnamefont {Kramer}},
  \bibinfo {author} {\bibfnamefont {M.}~\bibnamefont {MacCallum}}, \bibinfo
  {author} {\bibfnamefont {C.}~\bibnamefont {Hoenselaers}}, \ and\ \bibinfo
  {author} {\bibfnamefont {E.}~\bibnamefont {Herlt}},\ }\href {\doibase
  10.1017/CBO9780511535185} {\emph {\bibinfo {title} {Exact Solutions of
  Einstein's Field Equations}}},\ \bibinfo {edition} {2nd}\ ed.,\ Cambridge
  Monographs on Mathematical Physics\ (\bibinfo  {publisher} {Cambridge
  University Press},\ \bibinfo {year} {2003})\BibitemShut {NoStop}%
\bibitem [{\citenamefont {Kundt}\ and\ \citenamefont
  {Trumper}(1966)}]{Kundt:1966zz}%
  \BibitemOpen
  \bibfield  {author} {\bibinfo {author} {\bibfnamefont {W.}~\bibnamefont
  {Kundt}}\ and\ \bibinfo {author} {\bibfnamefont {M.}~\bibnamefont
  {Trumper}},\ }\bibfield  {title} {\enquote {\bibinfo {title} {{Orthogonal
  decomposition of axi-symmetric stationary spacetimes}},}\ }\href {\doibase
  10.1007/BF01325677} {\bibfield  {journal} {\bibinfo  {journal} {Z. Phys.}\
  }\textbf {\bibinfo {volume} {192}},\ \bibinfo {pages} {419--422} (\bibinfo
  {year} {1966})}\BibitemShut {NoStop}%
\bibitem [{\citenamefont {Carter}(1969)}]{Carter:1969zz}%
  \BibitemOpen
  \bibfield  {author} {\bibinfo {author} {\bibfnamefont {B.}~\bibnamefont
  {Carter}},\ }\bibfield  {title} {\enquote {\bibinfo {title} {{Killing
  horizons and orthogonally transitive groups in space-time}},}\ }\href
  {\doibase 10.1063/1.1664763} {\bibfield  {journal} {\bibinfo  {journal} {J.
  Math. Phys.}\ }\textbf {\bibinfo {volume} {10}},\ \bibinfo {pages} {70--81}
  (\bibinfo {year} {1969})}\BibitemShut {NoStop}%
\bibitem [{\citenamefont {Carter}(1973)}]{Carter:1973rla}%
  \BibitemOpen
  \bibfield  {author} {\bibinfo {author} {\bibfnamefont {B.}~\bibnamefont
  {Carter}},\ }\bibfield  {title} {\enquote {\bibinfo {title} {{Black holes
  equilibrium states}},}\ }in\ \href@noop {} {\emph {\bibinfo {booktitle} {{Les
  Houches Summer School of Theoretical Physics}: {Black Holes}}}}\ (\bibinfo
  {year} {1973})\ pp.\ \bibinfo {pages} {57--214}\BibitemShut {NoStop}%
\bibitem [{\citenamefont {Candelas}(1980)}]{Candelas:1980zt}%
  \BibitemOpen
  \bibfield  {author} {\bibinfo {author} {\bibfnamefont {P.}~\bibnamefont
  {Candelas}},\ }\bibfield  {title} {\enquote {\bibinfo {title} {{Vacuum
  Polarization in Schwarzschild Space-Time}},}\ }\href {\doibase
  10.1103/PhysRevD.21.2185} {\bibfield  {journal} {\bibinfo  {journal} {Phys.
  Rev. D}\ }\textbf {\bibinfo {volume} {21}},\ \bibinfo {pages} {2185--2202}
  (\bibinfo {year} {1980})}\BibitemShut {NoStop}%
\bibitem [{\citenamefont {Frolov}\ and\ \citenamefont
  {Novikov}(1998)}]{Frolov:1998wf}%
  \BibitemOpen
  \bibfield  {author} {\bibinfo {author} {\bibfnamefont {V.~P.}\ \bibnamefont
  {Frolov}}\ and\ \bibinfo {author} {\bibfnamefont {I.~D.}\ \bibnamefont
  {Novikov}},\ }\href {\doibase 10.1007/978-94-011-5139-9} {\emph {\bibinfo
  {title} {{Black hole physics: Basic concepts and new developments}}}}\
  (\bibinfo  {publisher} {Kluwer Acad. Publ.},\ \bibinfo {year}
  {1998})\BibitemShut {NoStop}%
\bibitem [{\citenamefont {Smarr}(1973)}]{Smarr:1973zz}%
  \BibitemOpen
  \bibfield  {author} {\bibinfo {author} {\bibfnamefont {L.}~\bibnamefont
  {Smarr}},\ }\bibfield  {title} {\enquote {\bibinfo {title} {{Surface Geometry
  of Charged Rotating Black Holes}},}\ }\href {\doibase 10.1103/PhysRevD.7.289}
  {\bibfield  {journal} {\bibinfo  {journal} {Phys. Rev. D}\ }\textbf {\bibinfo
  {volume} {7}},\ \bibinfo {pages} {289--295} (\bibinfo {year}
  {1973})}\BibitemShut {NoStop}%
\bibitem [{\citenamefont {Eguchi}\ \emph {et~al.}(1980)\citenamefont {Eguchi},
  \citenamefont {Gilkey},\ and\ \citenamefont {Hanson}}]{Eguchi:1980jx}%
  \BibitemOpen
  \bibfield  {author} {\bibinfo {author} {\bibfnamefont {T.}~\bibnamefont
  {Eguchi}}, \bibinfo {author} {\bibfnamefont {P.~B.}\ \bibnamefont {Gilkey}},
  \ and\ \bibinfo {author} {\bibfnamefont {A.~J.}\ \bibnamefont {Hanson}},\
  }\bibfield  {title} {\enquote {\bibinfo {title} {{Gravitation, Gauge Theories
  and Differential Geometry}},}\ }\href {\doibase 10.1016/0370-1573(80)90130-1}
  {\bibfield  {journal} {\bibinfo  {journal} {Phys. Rept.}\ }\textbf {\bibinfo
  {volume} {66}},\ \bibinfo {pages} {213} (\bibinfo {year} {1980})}\BibitemShut
  {NoStop}%
\bibitem [{\citenamefont {Miskovic}\ and\ \citenamefont
  {Olea}(2009)}]{Miskovic:2009bm}%
  \BibitemOpen
  \bibfield  {author} {\bibinfo {author} {\bibfnamefont {O.}~\bibnamefont
  {Miskovic}}\ and\ \bibinfo {author} {\bibfnamefont {R.}~\bibnamefont
  {Olea}},\ }\bibfield  {title} {\enquote {\bibinfo {title} {{Topological
  regularization and self-duality in four-dimensional anti-de Sitter
  gravity}},}\ }\href {\doibase 10.1103/PhysRevD.79.124020} {\bibfield
  {journal} {\bibinfo  {journal} {Phys. Rev. D}\ }\textbf {\bibinfo {volume}
  {79}},\ \bibinfo {pages} {124020} (\bibinfo {year} {2009})},\ \Eprint
  {http://arxiv.org/abs/0902.2082} {arXiv:0902.2082 [hep-th]} \BibitemShut
  {NoStop}%
\bibitem [{\citenamefont {Flachi}\ and\ \citenamefont
  {Fukushima}(2018)}]{Flachi:2017vlp}%
  \BibitemOpen
  \bibfield  {author} {\bibinfo {author} {\bibfnamefont {A.}~\bibnamefont
  {Flachi}}\ and\ \bibinfo {author} {\bibfnamefont {K.}~\bibnamefont
  {Fukushima}},\ }\bibfield  {title} {\enquote {\bibinfo {title} {{Chiral
  vortical effect with finite rotation, temperature, and curvature}},}\ }\href
  {\doibase 10.1103/PhysRevD.98.096011} {\bibfield  {journal} {\bibinfo
  {journal} {Phys. Rev. D}\ }\textbf {\bibinfo {volume} {98}},\ \bibinfo
  {pages} {096011} (\bibinfo {year} {2018})},\ \Eprint
  {http://arxiv.org/abs/1702.04753} {arXiv:1702.04753 [hep-th]} \BibitemShut
  {NoStop}%
\bibitem [{\citenamefont {Stone}\ and\ \citenamefont
  {Kim}(2018)}]{Stone:2018zel}%
  \BibitemOpen
  \bibfield  {author} {\bibinfo {author} {\bibfnamefont {M.}~\bibnamefont
  {Stone}}\ and\ \bibinfo {author} {\bibfnamefont {J.}~\bibnamefont {Kim}},\
  }\bibfield  {title} {\enquote {\bibinfo {title} {{Mixed Anomalies: Chiral
  Vortical Effect and the Sommerfeld Expansion}},}\ }\href {\doibase
  10.1103/PhysRevD.98.025012} {\bibfield  {journal} {\bibinfo  {journal} {Phys.
  Rev. D}\ }\textbf {\bibinfo {volume} {98}},\ \bibinfo {pages} {025012}
  (\bibinfo {year} {2018})},\ \Eprint {http://arxiv.org/abs/1804.08668}
  {arXiv:1804.08668 [cond-mat.mes-hall]} \BibitemShut {NoStop}%
\bibitem [{\citenamefont {Kraniotis}(2022)}]{Kraniotis:2021qah}%
  \BibitemOpen
  \bibfield  {author} {\bibinfo {author} {\bibfnamefont {G.~V.}\ \bibnamefont
  {Kraniotis}},\ }\bibfield  {title} {\enquote {\bibinfo {title} {{Curvature
  Invariants for accelerating Kerr-Newman black holes in (anti-)de Sitter
  spacetime}},}\ }\href {\doibase 10.1088/1361-6382/ac750a} {\bibfield
  {journal} {\bibinfo  {journal} {Class. Quant. Grav.}\ }\textbf {\bibinfo
  {volume} {39}},\ \bibinfo {pages} {145002} (\bibinfo {year} {2022})},\
  \Eprint {http://arxiv.org/abs/2112.01235} {arXiv:2112.01235 [gr-qc]}
  \BibitemShut {NoStop}%
\end{thebibliography}

%

\end{document}